\documentclass[preprints,review,accept,moreauthors,pdftex,10pt,a4paper]{mdpi}

\firstpage{1}
\articlenumber{x}
\doinum{10.3390/------}
\pubvolume{xx}
\pubyear{2019}
\copyrightyear{2019}
\history{Received: 27 March 2019; Accepted: 26 June 2019; Published: 1 July 2019}

\usepackage{bm}
\usepackage{amssymb}
\usepackage[utf8]{inputenc}

\Title{Halo Substructure Boosts to the Signatures of Dark Matter Annihilation}

\Author{Shin'ichiro Ando~$^{1,2,}$*\orcidA{}, Tomoaki Ishiyama~$^{3}$ and Nagisa Hiroshima~$^{4,5,6}$}

\AuthorNames{Shin'ichiro Ando, Tomoaki Ishiyama and Nagisa Hiroshima}

\address{%
$^{1}$ \quad GRAPPA Institute, Institute of Physics, University of Amsterdam, 1098 XH Amsterdam, The Netherlands\\
$^{2}$ \quad Kavli Institute for the Physics and Mathematics of the Universe (WPI), University of Tokyo, Kashiwa~277-8583,~Japan\\
$^{3}$ \quad Institute of Management and Information Technologies, Chiba University, Chiba 263-8522, Japan\\
$^{4}$ \quad Institute for Cosmic Ray Research, University of Tokyo, Kashiwa 277-8582, Japan\\
$^{5}$ \quad Institute of Particle and Nuclear Studies, High Energy Accelerator Research Organization (KEK), Tsukuba 305-0801, Japan\\
$^{6}$ \quad RIKEN Interdisciplinary Theoretical and Mathematical Sciences (iTHEMS),
Wako 351-0198, Japan}

\corres{Correspondence: s.ando@uva.nl}

\abstract{The presence of dark matter substructure will boost the signatures of dark matter annihilation. We review recent progress on estimates of this subhalo boost factor---a ratio of the luminosity from annihilation in the subhalos to that originating the smooth component---based on both numerical $N$-body simulations and semi-analytic modelings. Since subhalos of all the scales, ranging from the Earth mass (as expected, e.g., the supersymmetric neutralino, a prime candidate for cold dark matter) to galaxies or larger, give substantial contribution to the annihilation rate, it is essential to understand subhalo properties over a large dynamic range of more than twenty orders of magnitude in masses. Even though numerical simulations give the most accurate assessment in resolved regimes, extrapolating the subhalo properties down in sub-grid scales comes with great uncertainties---a straightforward extrapolation yields a very large amount of the subhalo boost factor of $\gtrsim$100 for galaxy-size halos. Physically motivated theoretical models based on analytic prescriptions such as the extended Press-Schechter formalism and tidal stripping modeling, which are well tested against the simulation results, predict a more modest boost of order unity for the galaxy-size halos. Giving an accurate assessment of the boost factor is essential for indirect dark matter searches and thus, having models calibrated at large ranges of host masses and redshifts, is strongly urged upon.}


\begin{document}

\section{Introduction}

One of the most popular candidates for dark matter is weakly interacting massive particles (WIMPs)~\cite{Bertone:2004pz, Bringmann:2012ez}.
{They are motivated by beyond-the-standard-model physics such as supersymmetry~\cite{Jungman:1995df} or universal extra-dimensions~\cite{Hooper:2007qk}, although the~non-discovery of new physics at the TeV scale with the Large Hadron Collider puts these models to serious test~\cite{Boveia:2018yeb}.  In~addition,} WIMPs can naturally
explain the relic dark matter density with thermal freezeout mechanisms, where {the WIMPs following the} weak-scale physics were in chemical equilibrium until freezeout---when the expansion of the Universe became faster than the annihilation rate~\cite{Steigman:2012nb}.
Since dark matter is often the lightest particle in an extended sector, it can self-annihilate only into the standard-model particles, which end up producing gamma rays, charged cosmic rays, and~neutrinos. 
Indirect detection of dark matter annihilation is therefore a direct test of the thermal freezeout of~WIMPs.

WIMPs are also {a} subcategory of cold dark matter (CDM), where they were nonrelativistic when structure formation started. In~the CDM framework, it is known that the structures form hierarchically, from~smaller to larger ones.
These virialized structures are referred to as halos and they are nearly spherically symmetric.
Typical size of the smallest structure is highly model dependent.
In the case of the supersymmetric neutralino that is one of the most popular WIMP candidates, the~smallest halos tend to be of the Earth mass, 10$^{-6} M_\odot$ but~with very large range of possible values of $\sim$10$^{-12}$--10$^{-3} M_\odot$~\cite{Hofmann:2001bi, Green:2003un, Loeb:2005pm, Bertschinger:2006nq, Profumo:2006bv, Bringmann:2009vf, Cornell:2012tb, Diamanti:2015kma}.
Smaller halos collapse at higher redshifts when the Universe was denser, and~hence they are of higher density.
A larger dark matter halo today contains lots of substructures (or subhalos) of all mass scales, which can go down to the Earth masses or even smaller and hence~denser.

Since the annihilation rate depends on the dark matter density squared (and $\langle \rho^2 \rangle \ge \langle\rho\rangle^2$), the~presence of the subhalos will \textit{boost} the gamma-ray signatures from dark matter annihilation.
This subhalo boost of dark matter annihilation, in~relation with the smallest-scale subhalos, has been a topic of interest for very many years~{\cite{Silk:1992bh,Bergstrom:1998jj, Bergstrom:1998zs, CalcaneoRoldan:2000yt, Tasitsiomi:2002vh, Stoehr:2003hf, Koushiappas:2003bn, Baltz:2006sv, Ando:2005hr, Oda:2005nv, Pieri:2005pg, Koushiappas:2006qq, Ando:2006cr, Pieri:2007ir, Diemand:2006ik, Berezinsky:2006qm, Lavalle:1900wn, SiegalGaskins:2008ge, Ando:2008br, Lee:2008fm, Kamionkowski:2008vw, Springel:2008by, Springel:2008cc, Ando:2009fp, Kamionkowski:2010mi, Pinzke:2011ek, Gao:2011rf, Fornasa:2012gu, Ng:2013xha, Sanchez-Conde:2013yxa, Ishiyama:2014uoa, Bartels:2015uba, Stref:2016uzb, Moline:2016pbm, Zavala:2015ura, Hiroshima:2018kfv}}.
{The main difficulty is the fact that subhalos of all the scales ranging from the Earth mass (or even smaller) to larger masses (a significant fraction of their host's mass) give a substantial contribution to the annihilation rate.}
Covering this very large dynamic range is challenging even with the state-of-the-art numerical simulations.
In simulations of Milky-Way-size halos ($10^{12}M_\odot$)~\cite{Diemand:2008in, Springel:2008cc, Stadel:2008pn}, one can resolve only down to $10^4$--$10^5 M_\odot$, and~there still remains more than ten orders of magnitude to~reach.

We will review recent progress on the subhalo contribution to dark matter annihilation.
{(See~also Reference~\cite{Berezinsky:2014wya} for a review on generic processes that subhalos undergo.)}
We first discuss approaches using the numerical $N$-body simulations and estimate of the annihilation boost factor by adoping the results and extrapolation down to very-small-mass ranges.
To complement the approach based on simulations, we then review an analytical approach.
In the CDM framework, fraction of halos that collapse is described with the Press-Schechter formalism~\cite{Press:1973iz} based on spherical or ellipsoidal collapse models.
This has been further extended to accommodate collapsed regions within larger halos (excursion set or extended Press-Schechter formalism~\cite{Bond:1990iw}), which can be applied to address statistics of halo substructure.
More recent literature suggests that the annihilation boost factor, defined as the luminosity due to subhalos divided by the host luminosity, is {modest, ranging from order of unity to a few tens for galaxy-size halos}~\cite{Kamionkowski:2008vw, Bartels:2015uba, Zavala:2015ura, Moline:2016pbm, Stref:2016uzb, Hiroshima:2018kfv}.
This relatively mild amount of the annihilation boost makes the prospect of indirect dark matter searches less promising compared with earlier more optimistic predictions~\cite{Springel:2008by, Pinzke:2011ek,Gao:2011rf,Anderson:2015dpc}.
{We note that our focus is mainly on subhalo boost factors in extragalactic halos. For~the subhalo boosts in the Galactic halo, on~the other hand, we need to assess the spatial distribution of the subhalos too. The~$N$-body simulations described in Section~\ref{sec:sim} can address this issue but~again are subject to resolution issues as well as the baryonic effect. See, for example,~Reference~\cite{Stref:2016uzb} for an alternative approach adopting analytical prescription.}

This review is organized as follows.
In Section~\ref{sec:Formulation: A simple case}, we introduce basic concepts of density profiles, mass functions, and~the annihilation boost factors of the subhalos, starting with simple formulations.
Here we make some simplifying assumptions, which are to be addressed in later sections.
In Section~\ref{sec:sim}, we summarize the progress from the numerical simulations for the subhalos and the annihilation boost factors.
Section~\ref{sec:Formulation: A more realistic case} presents more recent approaches based on realistic formulation than Section~\ref{sec:Formulation: A simple case}. 
{In Section~\ref{sub:Models based on structure formation and tidal evolution}, we first show new analytic models that predict the subhalo mass functions well in agreement with the results from the numerical simulations for various ranges of the host masses and redshifts, and~that the annihilation boost factors are on the order of unity even for cluster-size halos.
Then, we summarize other semi-analytic approaches for computing the annihilation boost factors, based on self-similarity (Section~\ref{sub:Models for self-similar subhalos}) and universal phase-space clustering (Section~\ref{sub:Universal clustering of dark matter in phase space}) of the subhalos.
We conclude the review in Section~\ref{sec:Conclusions}.}
Finally, for~convenience, we summarize fitting functions for the subhalo mass functions, and~annihilation boost factors that can be applicable to nearly arbitrary masses and redshifts in Appendix~\ref{app:Fitting formula}.

\section{Formulation}
\label{sec:Formulation: A simple case}

In this section, we introduce several important quantities such as density profiles, subhalo mass function, and~the annihilation boost factors.
This section is based on a simplified analytic model, which in several aspects are unrealistic but~sets the basis for the latter discussions according to numerical simulations (Section~\ref{sec:sim}) and more sophisticated semi-analytical models (Section~\ref{sec:Formulation: A more realistic case}).

\subsection{Subhalo Boost~Factor}

The rate of dark matter annihilation is proportional to dark matter density squared, $\rho_\chi^2$, where $\chi$ represents the dark matter particle.
In the presence of substructure, $\rho_\chi$ is divided into two terms:
\begin{equation}
    \rho_{\chi}(\bm x) = \rho_{\chi,{\rm sm}}(\bm x)+\rho_{\chi,{\rm sh}} (\bm x),
\end{equation}
where $\rho_{{\rm sm}}$ and $\rho_{{\rm sh}}$ represent smooth and subhalo components, respectively.
(In the following, we omit the subscript $\chi$.)
The volume average of the density squared in a host halo characterized by its virial mass $M$ and redshift $z$, which is the relevant quantity for the indirect dark matter researches, is therefore written as
\begin{equation}
    \langle\rho^2(\bm x)\rangle_{M,z}=\langle\rho_{\rm sm}^2(\bm x)\rangle_{M,z} + \langle\rho_{\rm sh}^2(\bm x)\rangle_{M,z} + 2\langle\rho_{\rm sm}(\bm x)\rho_{\rm sh}(\bm x)\rangle_{M,z}.
    \label{eq:dissection}
\end{equation}
We assume that the smooth component $\rho_{\rm sm}$ is characterized by the following Navarro-Frenk-White (NFW) profile~\cite{Navarro:1995iw, Navarro:1996gj}:
\begin{equation}
\rho_{\rm NFW}(r) = \frac{\rho_s}{(r/r_s)(1+r/r_s)^2},
\label{eq:NFW}
\end{equation}
where $\rho_s$ is a characteristic density and $r_s$ is a scale radius.
These parameters, $\rho_s$ and $r_s$, are evaluated such that the volume integral of $\rho_{\rm NFW}$ yields the total halo mass $M$, and~thus we have $\rho_{\rm sm}(r) = (1-f_{\rm sh}) \rho_{\rm NFW}(r)$, where $f_{\rm sh}$ is defined as the mass fraction in the subhalos.
The first term is then~simply
\begin{equation}
    \langle\rho_{\rm sm}^2(\bm x)\rangle_{M,z} = \frac{1}{V} \int d^3x \rho_{\rm sm}^2(r) =  \frac{4\pi(1-f_{\rm sh})^2}{3V}\rho_s^2 r_s^3\left[1-\frac{1}{(1+c_{\rm vir})^3}\right],
    \label{eq:rho_sm squared}
\end{equation}
where $V = 4\pi r_{\rm vir}^3/3$ is the volume of the host out to its virial radius $r_{\rm vir}$, $c_{\rm vir} \equiv r_{\rm vir}/r_s$ is the concentration parameter.
The parameters characterizing the host profile -- $\rho_s$, $r_s$, and~$c_{\rm vir}$ -- are all functions of $M$ and $z$.\footnote{We note, however, that the concentration $c_{\rm vir}$ has a scatter, which is often characterized by a log-normal distribution, whose mean $\bar c_{\rm vir}$ is the function of $M$ and $z$. We will include this in the latter sections.}

Next, we evaluate the second term of Equation~(\ref{eq:dissection}), $\langle\rho_{\rm sh}^2(\bm x)\rangle$.
We characterize each subhalo $i$ with the location of its center $\bm{x}_{i}$ and mass $m_i$.
Density due to all the subhalos at a coordinate $\bm{x}$ is written as a sum of the density profile around the \textit{seed} of each subhalo, that is,
\begin{equation}
\rho_{\rm sh}(\bm{x}) = \int dm' \int d^3x' \sum_i \delta_D(m'-m_i) \delta_D^3(\bm{x}'-\bm{x}_i)m'u_{\rm sh}(\bm{x}-\bm{x}'|m'),
\label{eq:rho sh}
\end{equation}
where $\delta_D^N$ is the $N$-dimensional Dirac delta function, and~$u_{\rm sh}(r|m)$ defines the density profile of the subhalo with mass $m$ and is normalized to one after the volume integral.\footnote{For the sake of simplicity for analytic expressions, we assume that the suhbalo mass is the only parameter characterizing its density profile. One can introduce many more parameters to make the model more realistic.}
We define the ensemble average of the product of these delta functions as
\begin{equation}
    \frac{dn_{\rm sh}(\bm x,m)}{dm} = \left\langle \sum_i \delta_D(m-m_i) \delta_D^3(\bm{x}-\bm{x}_i)\right\rangle,
\end{equation}
its volume integral over the host halo as
\begin{equation}
    \frac{dN_{\rm sh}}{dm} = \int d^3x \frac{dn_{\rm sh}(\bm x,m)}{dm},
\end{equation}
and call both $dn_{\rm sh}/dm$ and $dN_{\rm sh}/dm$ the subhalo mass function.
We also obtain the mass fraction in the subhalos as
\begin{equation}
    f_{\rm sh}(M,z) = \frac{1}{M}\int dm m\frac{dN_{\rm sh}}{dm}.
    \label{eq:fsh}
\end{equation}

By multiplying Equation~(\ref{eq:rho sh}) by itself and taking both the ensemble and the volume averages, we~have
\begin{eqnarray}
\langle\rho_{\rm sh}^2(\bm x)\rangle_{M,z} &\equiv& \frac{1}{V}\int d^3 x \langle\rho_{\rm sh}^2(\bm x)\rangle \nonumber\\
&=& \frac{1}{V}\int d^3 x\int dm'\int d^3 x' \int dm'' \int d^3 x'' 
m'u_{\rm sh}(\bm{x}-\bm{x}'|m')m''u_{\rm sh}(\bm{x}-\bm{x}''|m'')
\nonumber\\&&\times 
\left\langle \sum_{i}\delta_D(m'-m_i)\delta_D^3(\bm x'-\bm x_i)
\sum_j\delta_D(m''-m_j)\delta_D^3(\bm x''-\bm x_j)\right\rangle  \nonumber\\
&=& \frac{1}{V}\int d^3 x \int dm' \int d^3 x' \frac{dn_{\rm sh}(\bm x',m')}{dm'} m'^2 u_{\rm sh}^2(\bm x-\bm x'|m') \nonumber\\
&=& \frac{4\pi}{3V}\int dm \frac{dN_{\rm sh}}{dm}\rho_{s,{\rm sh}}^2r_{s,{\rm sh}}^3\left[1-\frac{1}{(1+c_{t,{\rm sh}})^3}\right],
\label{eq:rho_sh squared}
\end{eqnarray}
where at the last equality we adopted the NFW function for the subhalo density profile $m u_{\rm sh}(r|m)$ with the scale radius $r_{s,{\rm sh}}$, characteristic density $\rho_{s,{\rm sh}}$, and~tidal truncation radius $r_{t,{\rm sh}} \equiv c_{t,{\rm sh}} r_{s,{\rm sh}}$ beyond which the subhalo density abruptly decreases to zero.
At the third equality of Equation~(\ref{eq:rho_sh squared}), we ignored the term arising from $j \neq i$ as we evaluate the quantity at one point $\bm x$ and assume that subhalos do not overlap. We note, however, that such a term becomes relevant for obtaining the two-point correlation function, or~the power spectrum; see References~\cite{Ando:2006cr, Ando:2009fp} for more~details.

We define the subhalo boost factor as the ratio of the total luminosity from dark matter annihilation in the subhalos and that from the smooth component {\it in the case that there is no substructure}.
By~comparing Equations~(\ref{eq:rho_sm squared}) and (\ref{eq:rho_sh squared}), and~remembering that the luminosity is proportional to the volume integral of the density squared, the~boost factor is simply written as
\begin{equation}
    B_{\rm sh}(M,z) = \frac{1}{L_{\rm host, 0}(M,z)}\int dm\frac{dN_{\rm sh}(m|M,z)}{dm}L_{\rm sh}(m),
    \label{eq:boost simple}
\end{equation}
where the subscript $0$ shows that this is a quantity in the case of no subhalo contributions.
Equation~(\ref{eq:boost simple}) is also valid for any other spherically symmetric density profiles than the~NFW.

{Finally, we evaluate the last cross-correlation term in Equation~(\ref{eq:dissection}). See also References~\cite{2012CoPhC.183..656C, Nezri:2012tu, Ando:2013ff, Stref:2016uzb, Hutten:2016jko}.}
Following a similar procedure as in Equation~(\ref{eq:rho_sh squared}), we have
\begin{eqnarray}
2\langle\rho_{\rm sm}(\bm x)\rho_{\rm sh}(\bm x)\rangle_{M,z} &=& \frac{2}{V}\int d^3x\langle \rho_{\rm sm}(\bm x)\rho_{\rm sh}(\bm x)\rangle \nonumber\\
&=& \frac{2}{V}\int d^3x \rho_{\rm sm}(\bm x)\int dm' \int d^3x' m'u_{\rm sh}(\bm x-\bm x'|m')
\frac{dn_{\rm sh}(\bm x',m')}{dm'} \nonumber\\
&\approx& \frac{2}{V}\int d^3x\rho_{\rm sm}(\bm x)\int dm m\frac{dn_{\rm sh}(\bm x,m)}{dm},
\label{eq:cross}
\end{eqnarray}
where in the last equality, we first used the fact that the subhalo density profile is much more sharply peaked than their spatial distribution, and~take $dn_{\rm sh}/dm$ out of $\bm x'$ integration adopting $\bm x' \approx \bm x$ as its spatial variable.
Second, we performed volume integral for $u(\bm x-\bm x'|m)$ over $\bm x'$ variable, which simply returns one, to~reach the last expression of Equation~(\ref{eq:cross}).
Then we assume that the spatial distribution of the subhalos is independent of their masses:
\begin{equation}
    \frac{dn_{\rm sh}(\bm x,m)}{dm} = P_{\rm sh}(\bm x) \frac{dN_{\rm sh}(m)}{dm},
\end{equation} 
where $P_{\rm m sh}(\bm x) d^3x$ represents the probability of finding a subhalo in a volume element $d^3x$ around $\bm x$.
With this and Equation~(\ref{eq:fsh}), we have
\begin{equation}
    2\langle\rho_{\rm sm}(\bm x)\rho_{\rm sh}(\bm x)\rangle_{M,z} = \frac{2f_{\rm sh}M}{V}\int d^3x\rho_{\rm sm}(\bm x)P_{\rm sh}(\bm x).
    \label{eq:cross2}
\end{equation}
For simplicity, we assume that the subhalos are distributed following the smooth NFW component.
In~this case, we have $\rho_{\rm sm}(\bm x) = (1-f_{\rm sh})MP_{\rm sh}(\bm x)$, and~\begin{equation}
    2\langle\rho_{\rm sm}(\bm x)\rho_{\rm sh}(\bm x)\rangle_{M,z} = \frac{2f_{\rm sh}}{1-f_{\rm sh}}\langle\rho_{\rm sm}^2(\bm x)\rangle= \frac{8\pi f_{\rm sh}(1-f_{\rm sh})}{3V}\rho_s^2r_s^3\left[1-\frac{1}{(1+c_{\rm vir})^3}\right].
\end{equation}

The luminosity from the smooth component in the presence of the subhalos ($L_{\rm sm}$) is related to the host luminosity in the subhalos' absence via $L_{\rm sm} = (1-f_{\rm sh})^2L_{\rm host,0}$, because~the density in the smooth component gets depleted by a factor of $1-f_{\rm sh}$, if~there are subhalos.
Thus, the~total luminosity from both the smooth component and the subhalos are given by
\begin{eqnarray}
    L_{\rm total} &=& L_{\rm sm} + L_{\rm sh} + L_{\rm cross}
    \nonumber\\
    &=& \left[(1-f_{\rm sh})^2+B_{\rm sh}+2f_{\rm sh}(1-f_{\rm sh})\right]L_{\rm host,0}
    \nonumber\\
    &=& \left(1-f_{\rm sh}^2+B_{\rm sh}\right)L_{\rm host,0}
    \label{eq:Ltotal}
\end{eqnarray}
Often, $L_{\rm total}/L_{\rm host,0}$ is also referred to as the subhalo boost factor in the literature.
Note, however, that we have not included the effect of sub-subhalos (and beyond) yet in this formalism.
{In order to accommodate it, in~the right-hand side of Equation~(\ref{eq:boost simple}), we need to include the sub-subhalo boost to the subhalo luminosity $L_{\rm sh}$. Thus, we replace $L_{\rm sh}$ with $(1-f_{\rm ssh}^2+B_{\rm ssh})L_{\rm sh}$, where the subscript ``ssh'' represents the contribution from the sub-subhalos. If~the subhalo mass fraction $f_{\rm sh}$ and the boost factor $B_{\rm sh}$ depend only on the host mass, then one can assume $f_{\rm ssh}(m) = f_{\rm sh}(m)$ and $B_{\rm ssh}(m) = B_{\rm sh}(m)$, and~repeat the calculations in an iterative manner. See, however, Section~\ref{sub:Models based on structure formation and tidal evolution} for a more realistic~treatment.}

{\subsection{Characterization of Dark Matter~Halos}
\label{sub:Density profile of dark matter halos}}

We shall discuss the density profile of dark matter halos that are characterized by the virial radius $r_{\rm vir}$, the~scale radius $r_{s}$, and~the characteristic density $\rho_{s}$.
The halo is virialized when a mean density within a region reaches some critical value times the critical density of the Universe at that time: $\Delta_{\rm vir}(z)\rho_{c}(z)$, where $\rho_{c}(z) = \rho_{c,0} [\Omega_{m}(1+z)^3+\Omega_{\Lambda}]$, $\rho_{c,0} = 3 H_0^2/(8\pi G)$ is the present critical density, $H_0 = 100 h$~km~s$^{-1}$~Mpc$^{-1}$ is the Hubble constant,  $\Omega_m$ and $\Omega_{\Lambda}$ are the density parameters for matter and the cosmological constant, respectively.
In CDM cosmology with the cosmological constant, this critical value is given as~\cite{Bryan:1997dn}
\begin{equation}
    \Delta_{\rm vir}(z) = 18\pi^2 + 82 d(z) - 39 d^2(z),
\end{equation}
where $d(z) = \Omega_m(1+z)^3/[\Omega_m(1+z)^3+\Omega_\Lambda] - 1$.
Given the virial mass $M$ and the redshift $z$ of the halo of interest, $r_{\rm vir}$ is therefore obtained by solving
\begin{equation}
    M = \frac{4\pi}{3}\Delta_{\rm vir}(z)\rho_{c}(z) r_{\rm vir}^3.
\end{equation}
Alternatively, one can define $M_{200}$ and $r_{200}$ via
\begin{equation}
    M_{200} = \frac{4\pi}{3}200\rho_{c}(z) r_{200}^3.
\end{equation}
$M_{200}$ is often adopted to define halo masses in $N$-body~simulations.

The concentration parameter $c_{\rm vir} \equiv r_{\rm vir}/r_s$ (or $c_{200}\equiv r_{200}/r_s$) has been studied with numerical simulations and found to be a function of $M$ and $z$.
It follows a log-normal distribution with the mean of $\bar c_{\rm vir}(M,z)$ (e.g., Reference~\cite{Correa:2015dva}) and the standard deviation of $\sigma_{\log c} \approx 0.13$~\cite{Ishiyama:2011af}.
The mean $\bar c_{\rm vir}$ has been calibrated at both large (galaxies, clusters) and very small (of Earth-mass size) halos, and~found to decreases as a function of $M$ and $z$. 
Once $c_{\rm vir}$ is drawn from the distribution, it is used to obtain $r_s = r_{\rm vir}/c_{\rm vir}$.
Finally, $\rho_s$ is obtained through the condition of having mass $M$ within $r_{\rm vir}$:
\begin{eqnarray}
    M &=& \int_0^{r_{\rm vir}} dr 4\pi r^2 \rho(r) = 4\pi\rho_s r_s^3 f(c_{\rm vir}),\label{eq:enclosed mass}\\
    f(x) &=& \ln(1+x)-\frac{x}{1+x},
\end{eqnarray}
where the second equality of Equation~(\ref{eq:enclosed mass}) holds in the case of the NFW~profile.

In the case of the subhalos, the~procedures above cannot be adopted.
This is because they are subject to tidal effects from the host, which strip masses away from the subhalos.
However, the~regions well inside the scale radius $r_s$---because of strong self-gravity---is resilient against the tidal force and hence the annihilation rate hardly changes.
{These tidal processes, therefore, make the subhalos more concentrated and hence effectively brighter compared with the field halos of the same mass.}
{In many analytical studies in the literature~\cite{Diemand:2006ik, SiegalGaskins:2008ge, Anderhalden:2013wd, Sanchez-Conde:2013yxa}, however, the~effect of tidal stripping was ignored and the concentration-mass relation of the field halos was adopted, which resulted in underestimate of the annihilation boost factor.}
This has been pointed out by Reference \citet{Bartels:2015uba} and will be discussed in Section~\ref{sec:Formulation: A more realistic case} (see also References~\cite{Zavala:2015ura, Moline:2016pbm}).

\section{Estimates of Annihilation Boost with Numerical~Simulations}\label{sec:sim}

\begin{figure}[H]
\centering
\includegraphics[width=15cm]{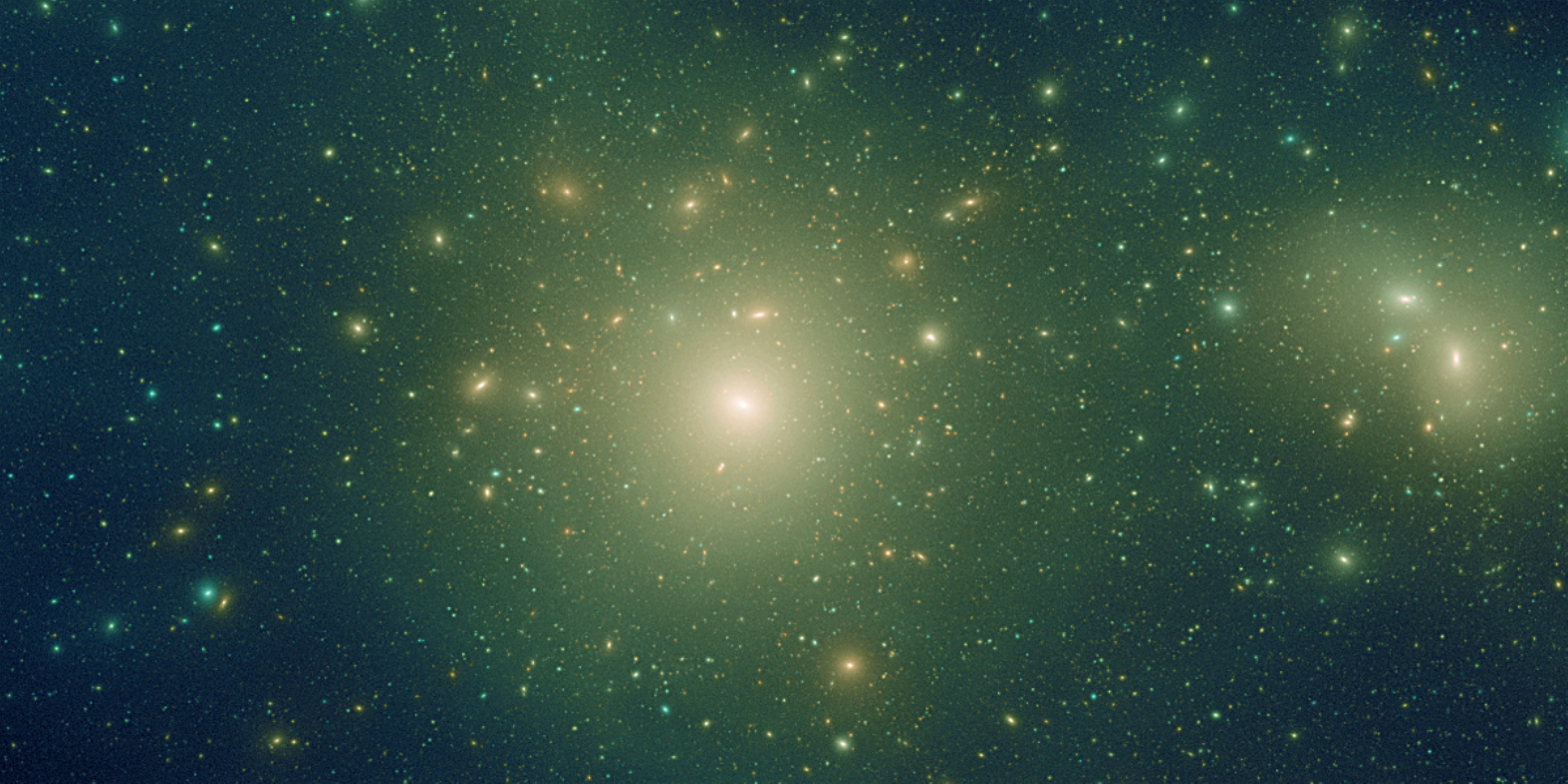}
\caption{
Dark matter distribution of a Milky-Way-size halo taken from a
high-resolution cosmological $N$-body simulation~\cite{2016ApJ...826....9I}.
}\label{fig:dm_distribution}
\end{figure}

In order to assess the annihilation boosts, one has to have reasonably good ideas on the density profiles $\rho(r)$, the~concentration-mass relation,\footnote{The concentration-mass relation is defined as the average concentration parameter as a function of halo mass.} and the subhalo mass function. 
Cosmological $N$-body simulations have been a powerful tool for probing all of them because
once a halo collapses from initial density fluctuations, it evolves
under a strongly nonlinear environment.  They have indeed demonstrated
that there are a large amount of surviving subhalos (see Figure~\ref{fig:dm_distribution}) in halos and halos have cuspy density~profiles.

\unskip

\subsection{Subhalo~Abundance}\label{subsec:subhalo_abundance}

Cosmological $N$-body simulations predict that there are many
surviving subhalos in host halos as a consequence of hierarchical structure
formation.  \citet{Klypin:1999uc} and \citet{Moore:1999nt} performed
high-resolution cosmological $N$-body simulations for the formation
and evolution of galaxy-scale halos.  They demonstrated that too many
subhalos existed in simulated halos in comparison with the number of
observed dwarf galaxies in the Local Group.  This discrepancy is known
as the ``missing satellite problem'' that has been investigated by a number of follow-up simulation studies (e.g., References~\cite{Diemand:2004kx,Reed:2004vg,
  Kase:2006wd}).
Even though it triggered many studies attempting to reduce small-scale structures by imposing other non-CDM candidates such as warm dark matter~\cite{Kamionkowski:1999vp} and self-interacting dark matter~\cite{Spergel:1999mh}, it is also possible to solve it with standard baryonic physics including early reionization~\cite{Susa:2004ji} and self-regulation of star formation in low-mass halos~\cite{Benson:2001at, Stoehr:2002ht,
  Kravtsov:2004cm, Okamoto:2009rz}.  Hence, it
is no longer regarded as a serious problem of the CDM~model.

These studies suggest a large number of ``dark satellites''
exist in halos, which do not contain optically visible components such as
gases and stars.  The~population of dark satellites is more abundant
in host halos than visible satellite galaxies and could enhance
annihilation boosts significantly.  To~estimate subhalo boosts to
annihilation signals accurately, understanding abundance of
subhalos as well as their density structure is~crucial.

A number of studies have calculated the subhalo mass function in halos using cosmological $N$-body simulations (e.g., References~\cite{Diemand:2006ik,
  Ishiyama:2008xe, Diemand:2008in, Springel:2008cc}), indicating 
that it obeys a power law
\begin{equation}
\frac{dN_{\rm sh}}{dm} \propto m^{-\alpha},
\label{eq:PLMF}
\end{equation}
where the slope $-\alpha$ ranges from $-2$ to $-1.8$, although~no
consensus has yet emerged. There is also a large halo-to-halo scatter
for the subhalo abundances~\cite{Ishiyama:2008xe, Mao:2015yua}. 
The subhalo abundance at a fixed mass halo depends on their accretion
history. Namely, it increases with the mass of halo and decreases with the 
halo concentration (e.g., References~\cite{Kravtsov:2003sg, Gao:2004au,
  Zentner:2004dq, vandenBosch:2004zs,Ishiyama:2008xe, Giocoli:2009ie,
  Gao:2010tn, Ishiyama:2011af, Mao:2015yua}).

Due to the limitation of currently available computational resources,
simulations cannot resolve the full hierarchy of subhalos from the
smallest to the most massive scales, which ranges more than twenty orders of magnitude in the mass. Even in the highest resolution
simulations for galaxy-scale halos, the~smallest resolved subhalo mass
is around $\sim$10$^5 M_{\odot}$~\cite{Diemand:2008in,Springel:2008cc}, which is still more than ten orders of magnitude more massive than that of the cutoff scale.  
To study subhalo boosts to annihilation signals, a~single-power-law subhalo mass function, Equation~(\ref{eq:PLMF}), is traditionally extrapolated beyond the~resolution. 

Another approach is to use some analytical
models (e.g., References~\cite{vandenBosch:2004zs, Giocoli:2007gf, Yang:2011rf, Jiang:2014nsa, Bartels:2015uba, Hiroshima:2018kfv}), which can shed light on the resolution issue. \citet{Hiroshima:2018kfv} developed  a model of the subhalo evolution calibrated with cosmological $N$-body simulations and found that the power-law index of the subhalo mass function is in
a rather narrow range between $-2$ and $-1.8$ with a vast range of subhalo
mass from $z = 0$ to 5.  
This picture is more or less consistent with the assumption of the subhalo mass function of the single power law.
More details on the analytic approach are discussed in the following~section.

Note that the annihilation boost factors strongly depend on the underlying
subhalo mass function~\cite{Sanchez-Conde:2013yxa, Ishiyama:2014uoa, Moline:2016pbm}.
Assuming that $\sim$10\% of the halo mass is within subhalos, the~difference of the boost factors in the Milky-Way-size halo could be as large as a factor of ten between the slope of $-\alpha=-2$ and $-1.9$~\cite{Sanchez-Conde:2013yxa}.  
More extensive simulations are needed to obtain the subhalo mass function in wide mass ranges and also to compare with analytic models~\cite{Hiroshima:2018kfv}.

\subsection{Density Profile of Dark Matter~Halos}\label{subsec:halo_prof}

By the end of the 1980's, it was already known in both
analytic~\cite{Gunn:1972sv} and numerical~\cite{1986Natur.322..329Q}
studies that the density profiles were described by power-law functions. Reference
\citet{Dubinski:1991bm} studied the density profiles of dark matter
halos using cosmological $N$-body simulations and argued that the
profiles were well described by a Hernquist
model~\cite{Hernquist:1990be}. 

Navarro~et~al.~\cite{Navarro:1995iw, Navarro:1996gj} simulated 
the structures of CDM halos systematically
with masses in the range of galaxy to rich cluster size.
They claimed that the radial density profile $\rho(r)$ could be described by a simple universal profile, Equation~(\ref{eq:NFW}), the~so-called NFW profile.
They also claimed that the shape of the profile was universal, independent of cosmological parameters, the~primordial power spectrum, and~the halo mass.  
Today, the~NFW profile has been extensively used to model halos analytically for various~purposes.

After the work of NFW, A number of subsequent studies (e.g., References~\cite{Fukushige:1996nr, Moore:1999gc}) performed simulations with better mass resolutions.
Whereas previous studies~\cite{Navarro:1995iw, Navarro:1996gj} used only $\sim$10,000 particles, they used $\sim$1,000,000 particles for a halo, and~found that the slope was steeper than $-1$.  
In the original results of the NFW, the~numerical two-body relaxation effects due to the small number of particles affected the structures of central regions and led to form a shallower cusp.  
Higher resolution simulations could resolve more inner structures of halos~\cite{Ghigna:1999sn, Jing:1999km, Fukushige:2000ar, 
Klypin:2000hk, Power:2002sw, Fukushige:2003xc, Hayashi:2003sj, Kazantzidis:2005su}.  
In most cases, the~slope of density became shallower as the radius went inward.  
A different approach was adopted by  \citet{Jing:1999ir}, who used the triaxial model for describing the central structures.
\citet{Moore:1999gc} and \citet{Diemand:2004wh} considered a more general profile,
\begin{equation}
\rho(r) = 
\frac{\rho_s}
{(r/r_{s})^{\gamma} \left[ 1+\left(r/r_{s} \right)^{\eta} \right]^
{(\beta-\gamma)/\eta}} .
\end{equation}
If $\beta=3$, $\gamma=1$, and~$\eta=1$, the~profile is the same as~NFW.

More recent studies~\cite{Diemand:2008in, Springel:2008cc, Stadel:2008pn} archived one of the highest resolution {dark matter only} simulations for galaxy-size halos with mass resolution better than $10^4M_{\odot}$.  
Their results are in agreement in that the density slope cannot be described by a single power law and the slope is around $-1$ at the radius $\sim$0.001$r_{200}$.  
Besides, \citet{Springel:2008cc} and \citet{Stadel:2008pn} fitted the density using the Einasto profile~\cite{Einasto1965} 
\begin{eqnarray}
\rho(r) = \rho_s \exp{\left\{-\frac{2}{\alpha_E}\left[\left(\frac{r}{r_{s}}\right)^{\alpha_E}-1 \right]\right\}},
\end{eqnarray}
where $\alpha_E$ is a free parameter.
Note that $r_s$ and $\rho_s$ are not the same parameters as those in Equation~(\ref{eq:NFW}).

Although we can obtain the density profile down to the radius $\sim$0.001$r_{200}$, the~result does not converge to a single power law.
In addition, the~physical origin of this flattening towards the center is not
understood at all.  However, the~importance of understanding the
central structures is increasing.  In~particular, if~we would like to
detect signals from dark matter annihilation, the~central structure of the dark matter halo is essential. 

The most important parameter to describe the halo profile is the concentration parameter.  
Assuming the universal NFW profile regardless of the halo mass, the~concentration-mass relation gives the annihilation rate as a function of the halo mass.  
Combined with assumed subhalo mass functions, they enable to estimate the annihilation boost factor.  
The concentration-mass relation of halos has been widely investigated and a number of fitting functions has been suggested~\cite{Bullock:1999he,Zhao:2003jf,Maccio:2006wpz, Neto:2007vq, Maccio:2008pcd, Zhao:2008wd, Klypin:2010qw, Prada:2011jf,  Ishiyama:2011af,Sanchez-Conde:2013yxa,Klypin:2014kpa,  Ishiyama:2014uoa,Correa:2015dva, Ludlow:2016ifl}.  
The concentration shows a weak dependence on the halo mass.  
The average concentration at fixed halo mass becomes smaller with increasing halo mass because the central density is tightly correlated with the cosmic density at the halo formation epoch, reflecting the hierarchical structure formation~\cite{Bullock:1999he,Wechsler:2001cs}.

Traditionally, the~concentration-mass relation has been calibrated with cosmological $N$-body simulations for relatively massive halos ($10^{10}M_\odot \lesssim M_{\rm 200} \lesssim 10^{15} M_{\odot}$).  
Because the mass dependence of the concentration is weak for these mass halos, it is found that a single power-law function, $c \propto M_{200}^{-\alpha_c}$, with~slope $\alpha_c$ in the range of 0.08 to 0.13, \
gives reasonable fits~\cite{Bullock:1999he, Maccio:2006wpz, Neto:2007vq, Maccio:2008pcd, Klypin:2010qw}.
However, the~dependence gradually becomes weaker toward less massive halos, and~a clear flattening emerges~\cite{Ishiyama:2011af,Sanchez-Conde:2013yxa,   Ishiyama:2014uoa, Correa:2015dva, Hellwing:2015upa, Ludlow:2016ifl, Pilipenko:2017iae}, ruling out single power-law concentration-mass relation 
for the full hierarchy of~subhalos.

{
These fitting functions are valid for the NFW density profile.
More generally, the~concentration can be defined independently
of the density profile and the subhalo mass as
(e.g., ~\cite{Diemand:2007qr,Springel:2008cc,Moline:2016pbm})
\begin{eqnarray}
c_{V} = \frac{\bar{\rho} (<r_{\rm max})}{\rho_{\rm c}}
= 2 \left( \frac{V_{\rm max}}{H_{0}r_{\rm max}} \right)^2,
\end{eqnarray}
where $r_{\rm max}$ is the radius at which the circular velocity reaches its maximum value $V_{\rm max}$.
This definition is also used to estimate the
annihilation boost factor (e.g., ~\cite{Moline:2016pbm}).
}

Even with the highest resolution simulations for galaxy-scale halos, the~smallest resolved subhalo mass is around $\sim$10$^5 M_{\odot}$~\cite{Diemand:2008in,Springel:2008cc}.
To estimate the annihilation boost from the full hierarchy of subhalos, we have to make some assumption of the concentration at unresolved scales, which has a significant impact on the result.  
One approach is extrapolating single power-law  fittings to the smallest scale beyond the mass range calibrated with simulations, although~the literature including the above cautions the risk of such extrapolations.  
With such extrapolations, the~concentration of the smallest halo can reach more than 100, substantially enhancing the annihilation boost. 
A number of studies have computed the concentration in such a manner and the resulting boost factor is a few hundreds for Milky-Way halos~\cite{Springel:2008by}, and~$\sim$1000 for cluster-scale halos~\cite{Pinzke:2011ek, Gao:2011rf, Anderson:2015dpc}. 

Another approach is adopting analytic models or fitting functions that can reproduce flattening of the concentration-mass relation 
{
(e.g.,~\cite{Prada:2011jf, Sanchez-Conde:2013yxa,Klypin:2014kpa, Correa:2015dva}).
}
In contrast to using the power-law extrapolation, the~resulting boost factor is rather modest, three to a few tens~\cite{Sanchez-Conde:2013yxa,Ishiyama:2014uoa,Stref:2016uzb} for Milky-Way halo, and~less than $\sim$100 for cluster-scale halos~\cite{Sanchez-Conde:2013yxa,Ishiyama:2014uoa,Anderson:2015dpc}.

The density profile at fixed halo mass shows a significant halo-to-halo scatter~\cite{Reed:2010gh}, possibly making a big impact on the annihilation signal.
Inferring from the cosmological Millennium simulation~\cite{Springel:2005nw}, the~effect of this non-universality on the annihilation flux is a factor of $\sim$3~\cite{Reed:2010gh}, which indicates that the uncertainty of the concentration-mass relation for low-mass halos has a more significant~effect.

These discussions are based on the universal density profile and the concentration-mass relation for field halos. 
There is a concern that whether or not we can apply the universal NFW profile for the full hierarchy of halos and subhalos beyond the range that cosmological simulations have been able to tackle.
We discuss this issue in Section~\ref{subsec:microhalo_prof}.
More importantly, we have to use the concentration-mass relation for subhalos, not field halos. 
We also discuss this issue in the following~section.

\subsection{Density Profile of Dark Matter~Subhalos}\label{subsec:subhalo_prof}

Density structures of subhalos are more challenging to be investigated
than field halos because it requires much higher resolutions.  Therefore, to~evaluate the subhalo contribution to the
annihilation signals, the~universal NFW profile and the
concentration-mass relation for field halos have been historically
assumed to be the same for subhalos as a first approximation, although~the underlying assumption is not well studied.  Complex physical
mechanisms relevant to subhalos could change their original density
profiles, such as the tidal effect from host halos, the~encounter with
other subhalos, and~denser environment than the~field.

\begin{figure}[H]
\centering
\includegraphics[width=12cm]{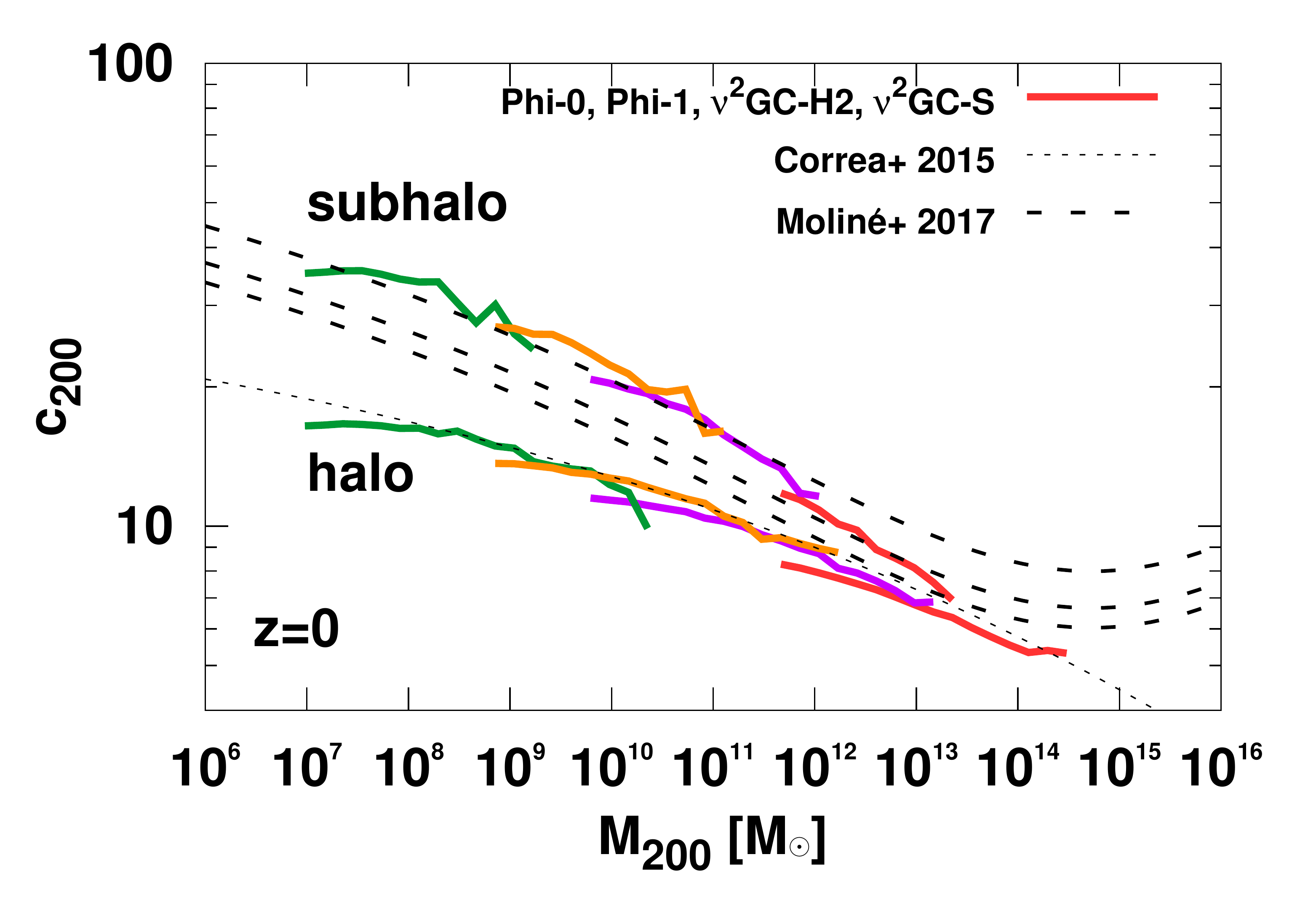}
\caption{  
The mass-concentraion relation of halos and subhalos at $z=0$, derived from 
high-resolution cosmological $N$-body simulations, Phi-0~\cite{2016ApJ...826....9I}, 
Phi-1, $\nu^2$GC-H2, and~$\nu^2$GC-S~\cite{Ishiyama:2014gla, Makiya:2015spa, Hiroshima:2018kfv} (green, orange, purple, and red, respectively).
Dashed and dotted curves are fitting formulae proposed by~\cite{Correa:2015dva,Moline:2016pbm}, respectively.
The dependence of the distance to host halo center gives three different dashed curves. 
}\label{fig:m-c200}
\end{figure}

Cosmological simulations have been suggesting that the density profile
of subhalos is cuspy in analogy with field halos. On~the other hand,
the average concentration of subhalos tend to be higher than those of
field halos (e.g.,~\cite{Bullock:1999he, Ghigna:1999sn,
  Diemand:2007qr, Springel:2008cc, Moline:2016pbm}).  For~example,
\citet{Bullock:1999he} showed that subhalos and halos tend to be more
concentrated in dense environments than in the field, and~the scatter of
concentrations is larger.  This result was taken into account to
estimate the gamma-ray flux from dark matter annihilation (e.g.,~\cite{Ullio:2002pj}).
\citet{Diemand:2007qr} showed that outer regions of subhalos tend to be
tidally stripped by host halos, which gives higher concentrations.
These results suggest that both earlier formation of halos/subhalos in
dense environments and tidal effect are responsible for the increased
concentration.  \citet{Pieri:2009je} derived the concentration-mass
relation of subhalos in Milky-Way-size halos by analyzing high
resolution cosmological simulations~\cite{Diemand:2008in,
  Springel:2008cc} and showed that it depends on the location of subhalos
relative to host halos.  Subhalos have considerably large
concentrations near the center than at the edge of host halos.
\citet{Moline:2016pbm} quantified the concentration of subhalos in
Milky-Way-size halos as a function of not only subhalo mass but
distance from host halo center, and~found a factor 2--3 enhancement of
the boost factor compared to the estimation that relied on the
concentration-mass relation of field halos (see also Figure~\ref{fig:m-c200}).

As shown in the literature in the above, higher concentrations of
subhalos than field halos could have a big impact on the annihilation
boost.  However, \citet{vandenBosch:2018tyt} argued that
subhalos even in state-of-the-art cosmological simulations 
suffer from excessive mass loss and artificial tidal disruption due to
inadequately large force softening (see also~\cite{Penarrubia:2010jk,
  vandenBosch:2017ynq}).  If~that is the case, it might be possible
that subhalos have larger concentrations than those ever considered.
These issues should be addressed by extremely high resolution
cosmological $N$-body simulations and analytic models
(e.g.,~\cite{Hiroshima:2018kfv}).

\subsection{Density Profile of Dark Matter Halos Near Cutoff~Scales}
\label{subsec:microhalo_prof}

In the CDM framework, smaller halos
collapse first, and~then they merge into more massive halos.  
Since the smallest halos contain no subhalos, their central structures might entirely differ from that observed in more massive halos.
If the dark matter particle is the lightest supersymmetric particle such as the neutralino, the~smallest halo mass is predicted to be around the Earth mass~\cite{Profumo:2006bv, Bringmann:2009vf, Diamanti:2015kma}.
Such~halos are sometimes referred to as ``microhalos.''

The density profiles of the microhalos have been investigated using cosmological $N$-body simulations~\cite{Diemand:2005vz, Ishiyama:2010es, Anderhalden:2013wd,   Ishiyama:2014uoa, Angulo:2016qof}.  \citet{Diemand:2005vz} simulated the formation of Earth-mass microhalos by means of cosmological $N$-body simulation.
They claimed that a single power law could describe the density profiles of microhalos, $\rho(r) \propto r^{-\gamma}$, with~a slope $\gamma$ in the range of 1.5 to 2.  As~a
consequence of such steep slope, most microhalos could not be completely destructed by the Galactic tide and encounters with stars, even in the Galactic center.  

\citet{Ishiyama:2010es} have performed $N$-body simulations with much higher resolution and showed that the density profile of microhalos had steeper cusps than the NFW profile.
The central density scales as $\rho(r) \propto r^{-1.5}$, which is supported by follow-up cosmological simulations~\cite{Anderhalden:2013wd, Ishiyama:2014uoa, Angulo:2016qof} and cold-collapse simulations~\cite{Ogiya:2017hbr}.
\citet{Ishiyama:2014uoa} has also shown that the cusp slope gradually becomes shallower with increasing halo mass.
Major merger of halos is responsible for the flattening, indicating that the process of violent relaxation plays a key role (see also~\cite{Ogiya:2016hyo,Angulo:2016qof}).  
Similar density structures are observed in recent simulations of ultracompact minihalos~\cite{Gosenca:2017ybi, Delos:2017thv, Delos:2018ueo} and warm dark matter~\cite{Polisensky:2015eya}.
{
The self-similar gravitational collapse models
(e.g.,~\cite{Gunn:1972sv, 1975ApJ...201..296G,Fillmore:1984wk,Bertschinger:1985pd,1995PhyU...38..687G})
can also give hints to understand the main physical origin of such steeper
cusps, because~the smallest halos do not contain smaller density
fluctuations by definition and collapse from initially overdense
patches.
}

Such microhalos with steep cusps can cause a significant effect on indirect dark matter searches.
\citet{Ishiyama:2010es} argued that the central parts of microhalos could survive against the encounters with stars except in the very Galactic center.
The nearest microhalos could be observable via gamma rays from dark matter annihilation, with~usually large proper motions of $\sim$0.2~deg~yr$^{-1}$, which are, however, stringently constrained with the diffuse gamma-ray backgrond~\cite{Ando:2008br}.  Gravitational perturbations to the 
millisecond pulsars might be detectable with future observations by pulsar timing arrays~\cite{Ishiyama:2010es,Baghram:2011is,Kashiyama:2018gsh, Dror:2019twh}.
\citet{Anderhalden:2013wd} have assumed a transition from the NFW to steeper cusps at scales corresponding to $\sim$100 times more massive than the cutoff and have found that such profiles can enhance moderately the annihilation boost of a Milky-Way-size halo by 5--12\%.
They also have found that concentrations of microhalos are consistent with a toy model proposed by \citet{Bullock:1999he}.

\begin{figure}[H]
\centering
\includegraphics[width=11.5cm]{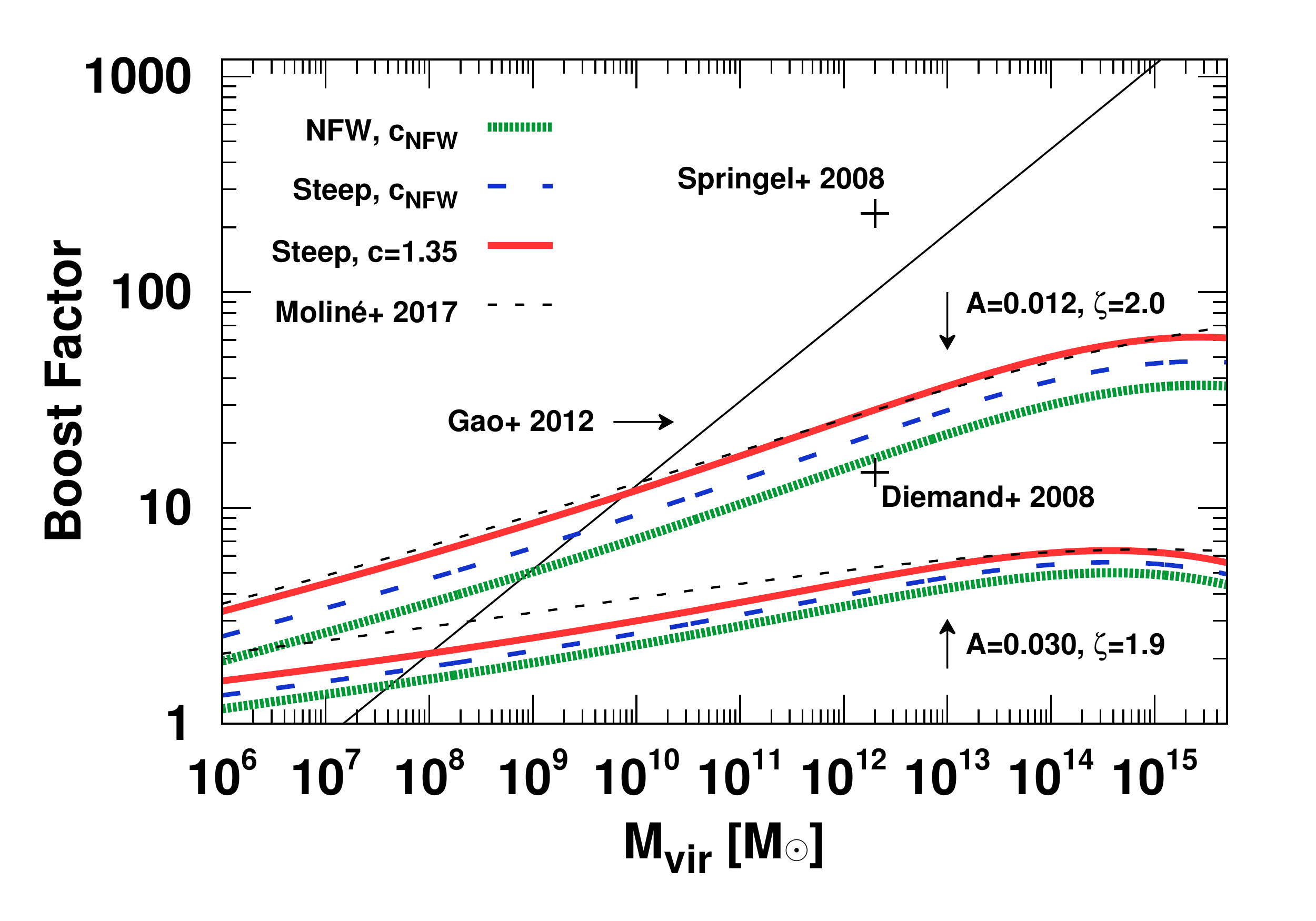}
\caption{Boost factor as a function of halo virial mass.
{The data used in six thick curves are taken from \citet{Ishiyama:2014uoa}. }
Two subhalo mass functions, 
$dn/dm=A/M_{\rm vir}(m/M_{\rm vir})^{-\xi}$, are used
{($A=0.012$, $\xi=2.0$, and~$A=0.030$, $\xi=1.9$~\cite{Sanchez-Conde:2013yxa, Moline:2016pbm})}.
Thick dotted curves are for the NFW profile, where the empirical concentration-mass relation of field halos~\cite{Sanchez-Conde:2013yxa} are assumed for the full hierarchy of subhalos.
Including the effect of steeper cusp of halos near the free streaming scale gives thick dashed curves. 
Besides, thick solid curves are results of incorporating the concentration of these halos derived from 
cosmological simulations~\cite{Ishiyama:2014uoa}.
For comparison, boost factors obtained in other studies are shown with 
thin dashed curves~\cite{Moline:2016pbm} (two subhalo mass functions are used), 
thin solid curves~\cite{Gao:2011rf}, and~crosses~\cite{Diemand:2008in, Springel:2008by}.}
\label{fig:mh_boost}
\end{figure}

\citet{Ishiyama:2014uoa} showed that the steeper inner cusps of halos in the smallest scale and near the cutoff scale could increase the annihilation rate of a Milky-Way-size halo by 12--67\%, compared with estimates adopting the universal NFW profile and an empirical concentration-mass relation~\cite{Sanchez-Conde:2013yxa} 
(see Figure~\ref{fig:mh_boost}).
The value, however, depends strongly on the adopted subhalo mass function and concentration model.  
They have found that concentrations near the free-streaming scale show little dependence on the halo mass and corresponding conventional NFW concentrations are 60--70, consistent with the picture that the mass dependence is gradually becoming weaker toward less massive halos (e.g.,~\cite{Ishiyama:2011af,Sanchez-Conde:2013yxa, Ishiyama:2014uoa,Correa:2015dva,Hellwing:2015upa,Ludlow:2016ifl}), ruling out a single power-law concentration-mass~relation.

As shown in the literature above, steep density cusps of halos near the free-streaming scale have an impact on the annihilation
boost.
However, these studies rely on the density structure seen in field halos, not subhalos.
It is also important to quantify the structures of subhalos near the free-streaming scale by larger simulations.
Another concern is that the cutoff in the matter power spectrum should suppress the number of subhalos near the free-streaming scale, which should weaken the annihilation signal.
However, the~shape of the mass function near the free-streaming scale is not understood well for the neutralino dark matter.
The structure of subhalos and the subhalo mass function near the free-streaming scale should be explored by larger volume cosmological $N$-body~simulations.

{
\section{Semi-Analytic~Approaches}
\label{sec:Formulation: A more realistic case}
\vspace{-6pt}

\subsection{Models Based on Structure Formation and Tidal~Evolution}
\label{sub:Models based on structure formation and tidal evolution}
}

In an analytical approach in Section~\ref{sec:Formulation: A simple case}, the~subhalo luminosity $L_{\rm sh}$ is characterized with the mass of the subhalo and the redshift of interest; see, for example,~Equation~(\ref{eq:boost simple}).
The mass and redshift, however, are not the only quantities that fully characterize the subhalo properties.
Indeed, they depend on the accretion history and mass loss after they fall onto their host halo,
that is, two subhalos that have the identical mass could have formed with different masses and accreted at different redshifts, evolved down to $z = 0$ reaching the same mass.
\citet{Bartels:2015uba} and \citet{Hiroshima:2018kfv} developed an analytical prescription to take these effects into account, which we follow in this~section.

{A subhalo is characterized} with its mass and redshift when it accreted onto its host, ($m_a$, $z_a$).
The concentration parameter $c_{a}$ is drawn from the log-normal distribution with mean $\bar c_{a}(m_{a},z_{a})$~\cite{Correa:2015dva} and $\sigma_{\log c} = 0.13$~\cite{Ishiyama:2011af}.
Since the subhalo was a \textit{field halo} when it just accreted, one can use the relations in Section~\ref{sub:Density profile of dark matter halos} to obtain $r_{s,a}$ and $\rho_{s,a}$ for the NFW~profile.

After the accretion, the~subhalos evolve by losing their mass through tidal forces.
The mass-loss rate is typically characterized by a dynamical timascale at the redshift $z$,
\begin{equation}
\tau_{\rm dyn}(z) =1.628h^{-1}\,{\rm Gyr}\left[\frac{\Delta_{\rm vir}(z)}{178}\right]^{1/2}\left[\frac{H(z)}{H_0}\right]^{-1},
\end{equation}
as follows~\cite{Jiang:2014nsa}:
\begin{equation}
    \dot m(z) = -A\frac{m(z)}{\tau_{\rm dyn}(z)}\left[\frac{m(z)}{M(z)}\right]^\zeta,
    \label{eq:mdot}
\end{equation}
where $H(z) = H_0 [\Omega_m(1+z)^3+\Omega_\Lambda]^{1/2}$, $m(z)$ and $M(z)$ are the subhalo and host-halo masses at $z$, respectively.
Following \citet{Jiang:2014nsa}, \citet{Hiroshima:2018kfv} adopted simple Monte Carlo simulations to estimate $\dot m$ based on the assumption that the subhalo loses all the masses beyond its tidal radius in one complete orbit at its peri-center passage.
While \citet{Jiang:2014nsa} found $A = 0.81$ and $\zeta = 0.04$, \citet{Hiroshima:2018kfv} extended the mass and redshift ranges of applicability and found that these parameters are weakly dependent on both $M$ and $z$:
\begin{eqnarray}
\log A &=& \left\{-0.0003\log\left[\frac{M(z)}{M_\odot}\right]+0.02\right\}z+0.011\log\left[\frac{M(z)}{M_\odot}\right]-0.354, \\
\zeta &=& \left\{0.00012\log\left[\frac{M(z)}{M_\odot}\right]+0.0033\right\}z+0.0011\log\left[\frac{M(z)}{M_\odot}\right]+0.026.
\end{eqnarray}
{One can} solve Equation~(\ref{eq:mdot}) to obtain the subhalo mass at a redshift of interest $z$, $m(z)$, with~a boundary condition of $m(z_a) = m_a$.
For the evolution of the host, $M(z)$, {\citet{Hiroshima:2018kfv}} adopted a fitting formula given by \citet{Correa:2014xma}.

{The subhalo density profile after accretion is also well described with the NFW profile with a sharp truncation at $r_t$:
\begin{equation}
\rho(r) = \left\{
\begin{array}{lcc}
\rho_s r_s^3/[r(r+r_s)]^2, & {\rm for} & r<r_t, \\
0, & {\rm for} & r\geq r_t.
\end{array}\right.
\end{equation}}
This is indeed a good approximation found in the simulations~\cite{Springel:2008cc}.
In addition to $r_t$, \mbox{\citet{Penarrubia:2010jk}}~found that the internal structure changes.
If the inner profile is $\propto r^{-1}$ just like NFW, the~maximum circular velocty $V_{\rm max}$ and its corresponding radius $r_{\rm max}$ evolve as
\begin{eqnarray}
\frac{V_{{\rm max}}(z)}{V_{{\rm max},a}}&=&\frac{2^{0.4}[m(z)/m_a]^{0.3}}{[1+m(z)/m_a]^{0.4}}, \\
\frac{r_{\rm max}(z)}{r_{\rm max,0}} &=& \frac{2^{-0.3}[m(z)/m_a]^{0.4}}{[1+m(z)/m_a]^{-0.3}},
\end{eqnarray}
respectively.
After computing $V_{\rm max}$ and $r_{\rm max}$ at $z$, one can convert them to $\rho_s$ and $r_s$ through
\begin{eqnarray}
r_s &=& \frac{r_{\rm max}}{2.163},\\
\rho_s &=& \frac{4.625}{4\pi G}\left(\frac{V_{\rm max}}{r_s}\right)^2,
\end{eqnarray}
which are valid for the NFW profile.
Finally by solving the condition
\begin{equation}
    m(z) = \int_0^{r_t}dr 4\pi r^2\rho(r) = 4\pi \rho_s r_s^3 f(r_t/r_s),
\end{equation}
{the truncation radius $r_t$ is obtained.} 
{\citet{Hiroshima:2018kfv} omitted subhalos with $r_t < 0.77 r_s$ from the subsequent calculations assuming that they were tidally disrupted~\cite{Hayashi:2002qv}. This criterion, however, might be a numerical artifact~\cite{vandenBosch:2017ynq}. Either case, \citet{Hiroshima:2018kfv} checked that whether one
implements this condition or not did not have impact on the results of, for example,~subhalo mass functions.}

Thus, given $(m_a, z_a, c_a)$, one can obtain all the subhalo parameters after the evolution, $(m, r_s, \rho_s, r_t)$, in~a deterministic manner.
The differential number of subhalos accreted onto a host with a mass $m_a$ and at redshift $z_a$, $d^2N_{\rm sh}/(dm_a dz_a)$, is given by the excursion set or the extended Press-Schechter formalism~\cite{Bond:1990iw}.
Especially \citet{Yang:2011rf} obtained analytical formulation for the distribution that provides good fit to the numerical simulation data over a large range of $m/M$ and $z$.
\mbox{{\citet{Hiroshima:2018kfv} adopted}} their model~III.

\begin{figure}[H]
\centering
\includegraphics[width=7.5cm]{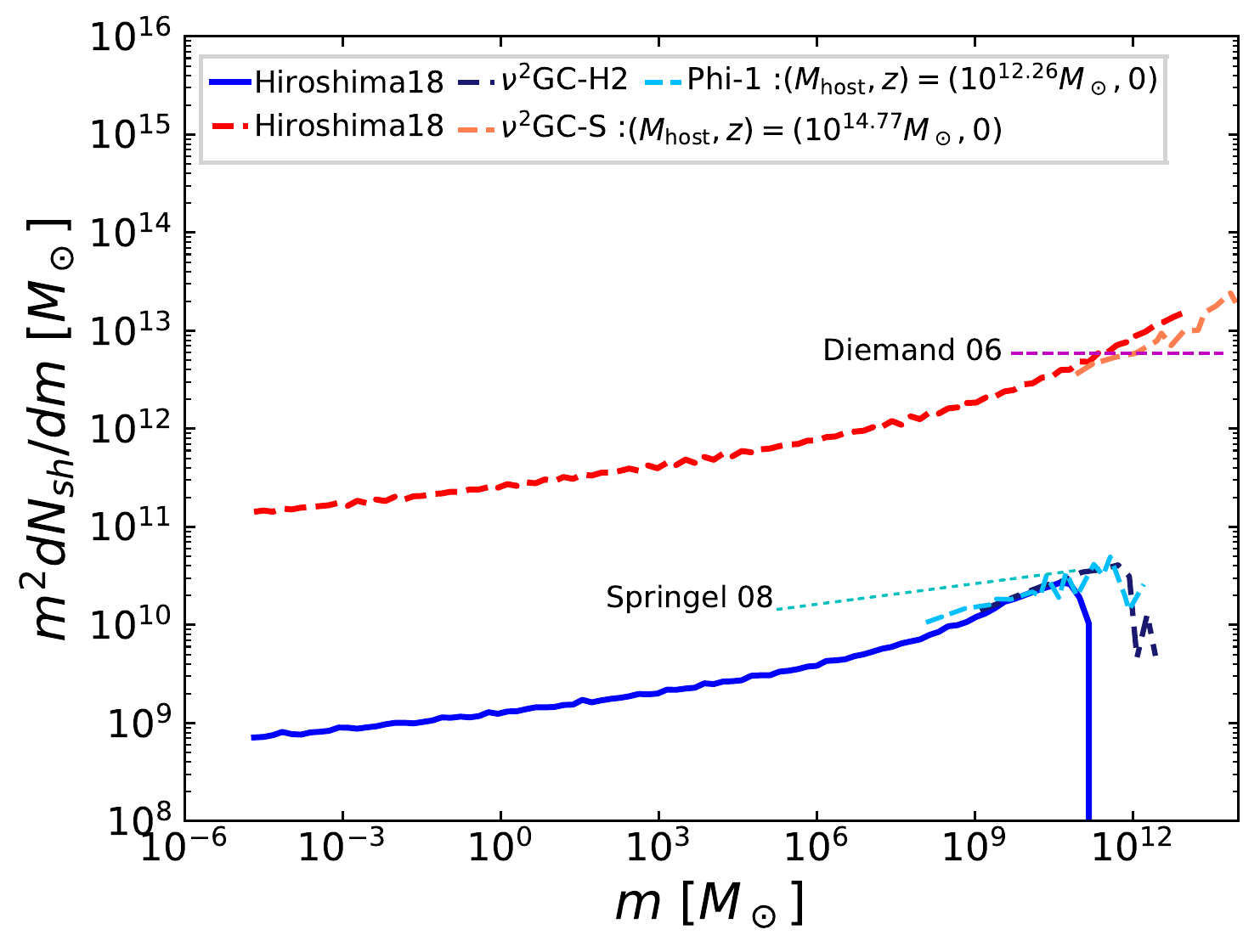}
\includegraphics[width=7.5cm]{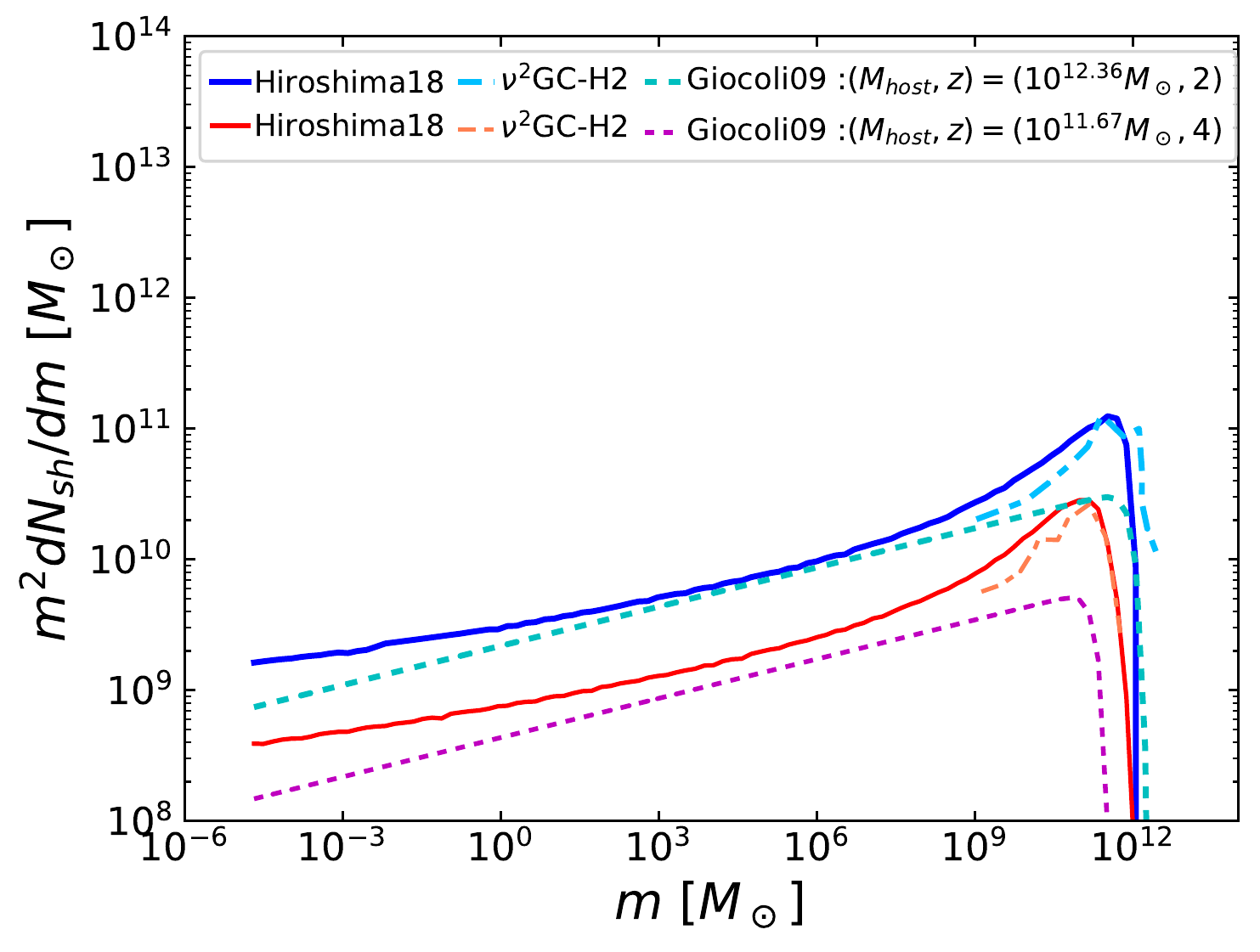}
\includegraphics[width=7.5cm]{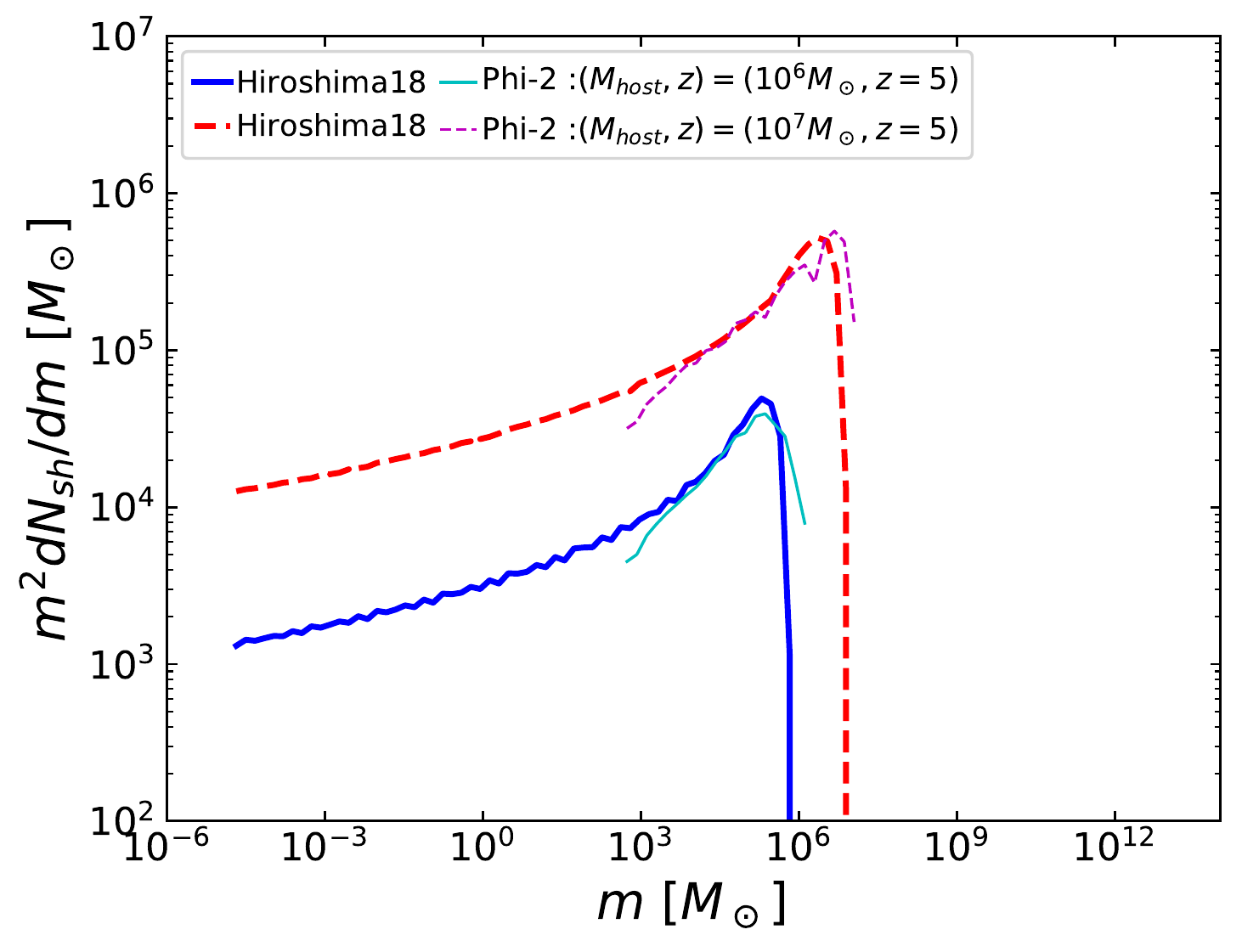}
\includegraphics[width=7.5cm]{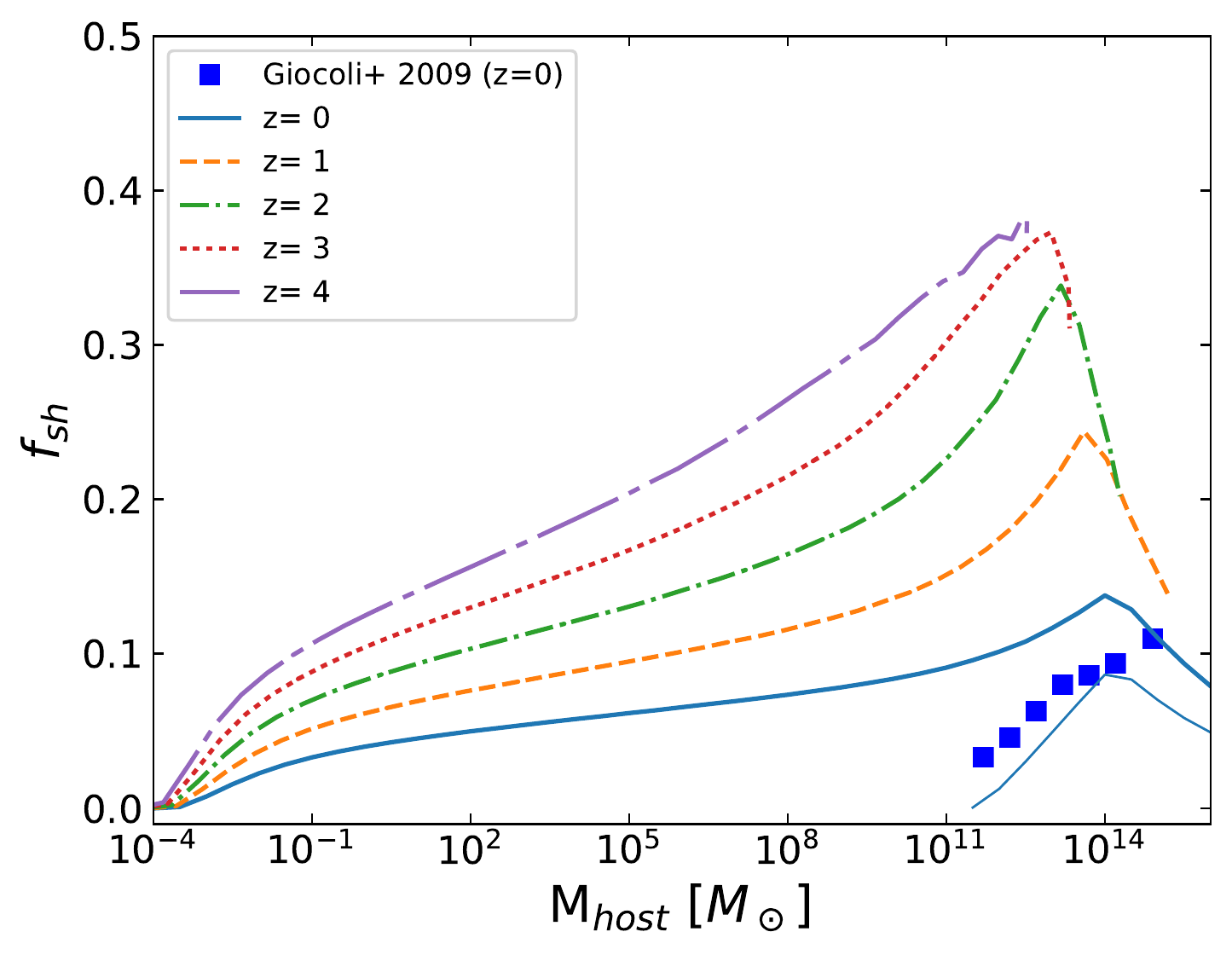}
\caption{Subhalo mass function for galaxy ($M_{200} = 1.8\times 10^{12}M_\odot$) and cluster ($M_{200} = 5.9\times 10^{14}M_\odot$) halos at $z = 0$ (top left), halos with $2.3\times 10^{12}M_\odot$ at $z = 2$ and $4.7\times 10^{11}M_\odot$ at $z=4$ both of which would evolve to $M_{200} = 10^{13}M_\odot$ at $z = 0$ (top right) and~smaller halos of $M_{200} = 10^6 M_\odot$ and $10^7 M_\odot$ at $z = 5$ (bottom left). {Results of the analytic models by \citet{Hiroshima:2018kfv} are compared with those from the numerical simulations of similar halos and other fitting functions:} \citet{Springel:2008cc}, \citet{Diemand:2006ey} (top left), $\nu^2$GC H2~\cite{Ishiyama:2014gla, Makiya:2015spa}, \citet{Giocoli:2009ie} (top right) and~Phi-2~(Ishiyama~et~al., in~preparation; bottom left). The~bottom right panel shows the subhalo mass fraction $f_{\rm sh}$ as a function of the host mass $M_{200}$ for various values of redshift $z$. The~thin solid curve is for $z = 0$ but with lower mass threshold of $1.73 \times 10^{10} h^{-1} M_\odot$ to be compared with \citet{Giocoli:2009ie} results shown as the~squares.}
\label{fig:mass function}
\end{figure}

The subhalo mass function is obtained as
\begin{equation}
    \frac{dN_{\rm sh}(m|M,z)}{dm} = \int dm_a \int dz_a \frac{d^2N_{\rm sh}}{dm_adz_a} \int dc_a P(c_a|m_a,z_a) \delta(m-m(z|m_a,z_a,c_a)),
    \label{eq:dNdm}
\end{equation}
where $P(c_a|m_a,z_a)$ is the probability distrbution for $c_a$ given $m_a$ and $z_a$, for~which \mbox{{\citet{Hiroshima:2018kfv} adopted}} the log-normal distribution with the mean $\bar c_a(m_a,z_a)$~\cite{Correa:2015dva} and the standard deviation $\sigma_{\log c_a} = 0.13$~\cite{Ishiyama:2011af}.
We show the subhalo mass functions obtained with Equation~(\ref{eq:dNdm}) for various values of $M$ and $z$ in Figure~\ref{fig:mass function}, where comparison is made with simulation results of similar host halos.
{
Halos and subhalos formed in these simulations were identified with ROCKSTAR phase space halo finder~\cite{Behroozi:2011ju}. The~bound mass is used as the subhalo mass, which nearly
corresponds to the tidal mass~\cite{Moline:2016pbm}.
}
For all these halos, one can see remarkable agreement between the analytic model and the corresponding simulation results in resolved regimes.
Successfully reproducing behaviors at resolved regimes, this analytic model is able to make reliable predictions of the subhalo mass functions below resolutions of the numerical simulations, without~relying on extrapolating a single power-law functions, from~which most of the previous studies in the literature had to suffer.
The subhalo mass fraction is then obtained as
\begin{eqnarray}
f_{\rm sh}(M,z) &=& \frac{1}{M}\int dm m\frac{dN_{\rm sh}(m|M,z)}{dm} 
\nonumber\\
&=& \frac{1}{M}\int dm_a \int dz_a \frac{d^2N_{\rm sh}}{dm_adz_a} \int dc_a P(c_a|m_a,z_a)m(z|m_a,z_a,c_a),
\label{eq:fsh2}
\end{eqnarray}
and is shown in Figure~\ref{fig:mass function} (bottom right) for various values of redshifts.
{The subhalo mass fraction is found to increase as a function of $M$ and $z$.}
At higher redshifts, since there is shorter time for the subhalos to experience tidal mass loss, $f_{\rm sh}$ is larger.
Again, {a good agreement in $f_{\rm sh}$ is found} between the analytic model and the simulation results by \citet{Giocoli:2009ie}.

The annihilation boost factor is then
\begin{equation}
    B_{\rm sh}^{(0)}(M,z) = \frac{1}{L_{\rm host,0}(M,z)}\int dm_a \int dz_a \frac{d^2N_{\rm sh}}{dm_adz_a} \int dc_a P(c_a|m_a,z_a) L_{\rm sh}^{(0)}(m_a,z_a,c_a|M,z),
    \label{eq:boost0}
\end{equation}
which is to be compared with Equation~(\ref{eq:boost simple}) that was derived with a simpler (and unrealistic) discussion.
The superscript (0) represents the quantity \textit{in the absense of sub-subhalos and beyond}.
The subhalo luminosity, $L_{\rm sh}^{(0)}(m_a,z_a,c_a|M,z)$, is proportional to the volume integral of density squared $\rho_{\rm sh}^2(r)$ out to the truncation radius,
\begin{eqnarray}
L_{\rm sh}^{(0)}(m_a,z_a,c_a|M,z) \propto \int d^3x\rho_{\rm sh}^2(\bm x) = \frac{4\pi}{3}\rho_{s,{\rm sh}}^2r_{s,{\rm sh}}^3\left[1-\frac{1}{(1+r_{t,{\rm sh}}/r_{s,{\rm sh}})^3}\right],
\label{eq:Lsh0}
\end{eqnarray}
where $\rho_{s,{\rm sh}}$, $r_{s,{\rm sh}}$ and~$r_{t,{\rm sh}}$ are functions of $(m_a,z_a,c_a)$ as well as $M$ and $z$.

{Then, the~effect of sub$^{n}$-subhalos (for $n\ge 1$) can be estimated} iteratively.
At $n$th iteraction, when a subhalo accreted onto its host at $z_a$ with $m_a$, {it is assigned} a sub-subhalo boost factor $B_{\rm sh}^{(n-1)}(m_a,z_a)$.
After the accretion, the~outer region of the subhalo is stripped away by the tidal force and~thus all the sub-subhalos within this stripped region will disappear, reducing the sub-subhalo boost accordingly.
{\citet{Hiroshima:2018kfv} assumed that the sub-subhalos were} distributed within the subhalo following $n_{\rm ssh}(r) \propto [1+(r/r_s)^2]^{-3/2}$.
The luminosity due to sub-subhalos within a radius $r$ is therefore proportional to their enclosed number
\begin{equation}
    N_{\rm ssh}(<r|r_s) = \int_0^r dr' 4\pi r'^2 n_{\rm ssh}(r') \propto r_s^3 \left[\sinh^{-1}\left(\frac{r}{r_s}\right)-\frac{r}{\sqrt{r^2+r_s^2}}\right],
    \label{eq:Nssh}
\end{equation}
and it gets suppressed by a factor of $N_{\rm ssh}(<r_t|r_s) / N_{\rm ssh}(<r_{\rm vir}|r_{s,a})$ due to the tidal stripping.\footnote{We note that in estimating the effect of sub$^n$-subhalos in the boost factors, Reference \citet{Hiroshima:2018kfv} ignored the changes of $\rho_s$ and $r_s$ and~hence did not include the factor of $r_s^3$ in Equation~(\ref{eq:Nssh}) and $\rho_s r_s^3$ in Equation~(\ref{eq:Lsh0enc}). In~addition, in~Equation~(\ref{eq:boost}), they multiplied $L_{\rm sh}^{(0)}$ by a factor of $1+B_{\rm ssh}^{(n)}$ instead of $1-f_{\rm ssh}^2+B_{\rm ssh}^{(n)}$. We correct for all these effects in this review.}
The luminosity due to the smooth component also decreases as $L_{\rm sh}^{(0)}(<r_t|\rho_s,r_s)/L_{\rm sh}^{(0)}(<r_{\rm vir}|\rho_{s,a},r_{s,a})$, where
\begin{equation}
     L_{\rm sh}^{(0)}(<r|\rho_s,r_s) \propto \rho_s^2r_s^3\left[1-\frac{1}{(1+r/r_s)^3}\right].
     \label{eq:Lsh0enc}
\end{equation}
Thus the sub-subhalo boost after the $n$th iteration, $B_{\rm ssh}^{(n)}$ is obtained by
\begin{equation}
    B_{\rm ssh}^{(n)}(m_a,z_a,c_a|M,z) = B_{\rm sh}^{(n-1)}(m_a,z_a)\frac{N_{\rm ssh}(<r_t|r_s)/N_{\rm ssh}(<r_{\rm vir}|r_{s,a})}{L_{\rm sh}^{(0)}(<r_t|\rho_s,r_s)/L_{\rm sh}^{(0)}(<r_{\rm vir}|\rho_{s,a},r_{s,a})}.
\end{equation}
Similarly, the~sub-subhalo mass fraction $f_{\rm ssh}$ is obtained by
\begin{equation}
    f_{\rm ssh}(m_a,z_a,c_a|M,z) = f_{\rm sh}(m_a,z_a)\frac{N_{\rm ssh}(<r_t|r_s)/N_{\rm ssh}(<r_{\rm vir}|r_{s,a})}{m_{\rm sm}(<r_t|\rho_s,r_s)/m_{\rm sm}(<r_{\rm vir}|\rho_{s,a},r_{s,a})},
\end{equation}
where $f_{\rm sh}(m_a,z_a)$ is obtained with Equations~(\ref{eq:fsh2}) and~$m_{\rm sm}(<r|\rho_s,r_s) \propto \rho_{s}r_{s}^3f(r/r_s)$ is the enclosed mass within $r$ of the smooth component of the subhalo.
The subhalo boost factor after $n$th iteration is obtained with Equation~(\ref{eq:boost0}) by replacing $L_{\rm sh}^{(0)}$ with $[1-f_{\rm ssh}^2 + B_{\rm ssh}^{(n)}]L_{\rm sh}^{(0)}$ [see discussions below Equation~(\ref{eq:Ltotal})]:
\begin{eqnarray}
    B_{\rm sh}^{(n)}(M,z) &=& \frac{1}{L_{\rm host,0}(M,z)}\int dm_a \int dz_a \frac{d^2N_{\rm sh}}{dm_adz_a} \int dc_a P(c_a|m_a,z_a)
    \nonumber\\&&{}\times
    \left[1-f_{\rm ssh}^2(m_a,z_a,c_a|M,z) + B_{\rm ssh}^{(n)}(m_a,z_a,c_a|M,z)\right] L_{\rm sh}^{(0)}(m_a,z_a,c_a|M,z).
    \label{eq:boost}
\end{eqnarray}
{The host luminosity in the absence of the subhalos $L_{\rm host,0}(M,z)$ is defined by} marginalizing over the concentration parameter $c_{\rm vir}$:
\begin{equation}
    L_{\rm host,0}(M,z) \propto \frac{4\pi}{3} \int dc_{\rm vir}P(c_{\rm vir}|M,z)\rho_s^2(M,z,c_{\rm vir})r_s^3(M,z,c_{\rm vir})\left[1-\frac{1}{(1+c_{\rm vir})^3}\right].
\end{equation}
with the log-normal distribution $P(c_{\rm vir}|M,z)$.

\begin{figure}[H]
    \centering
     \includegraphics[width=7.5cm]{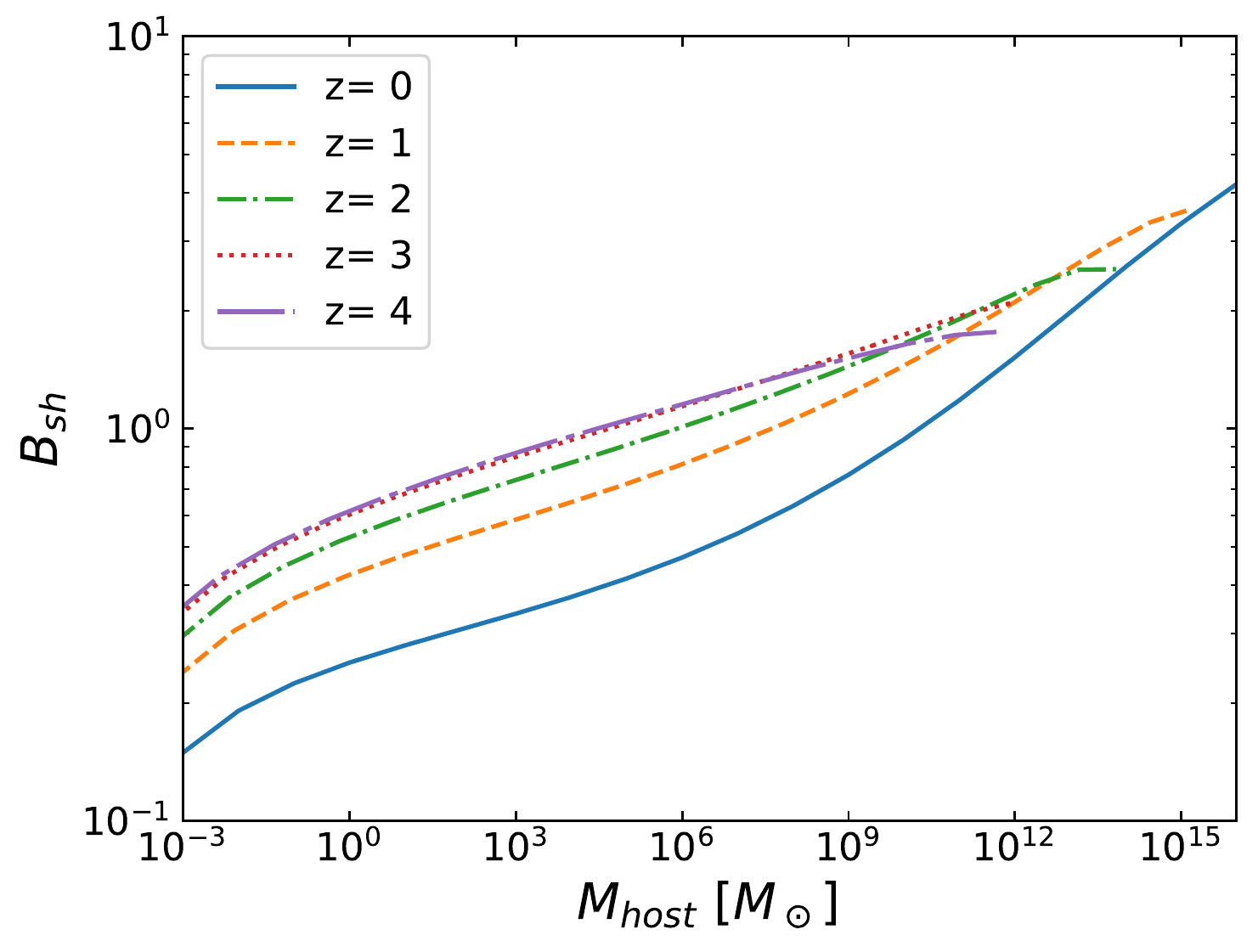}
     \includegraphics[width=7.5cm]{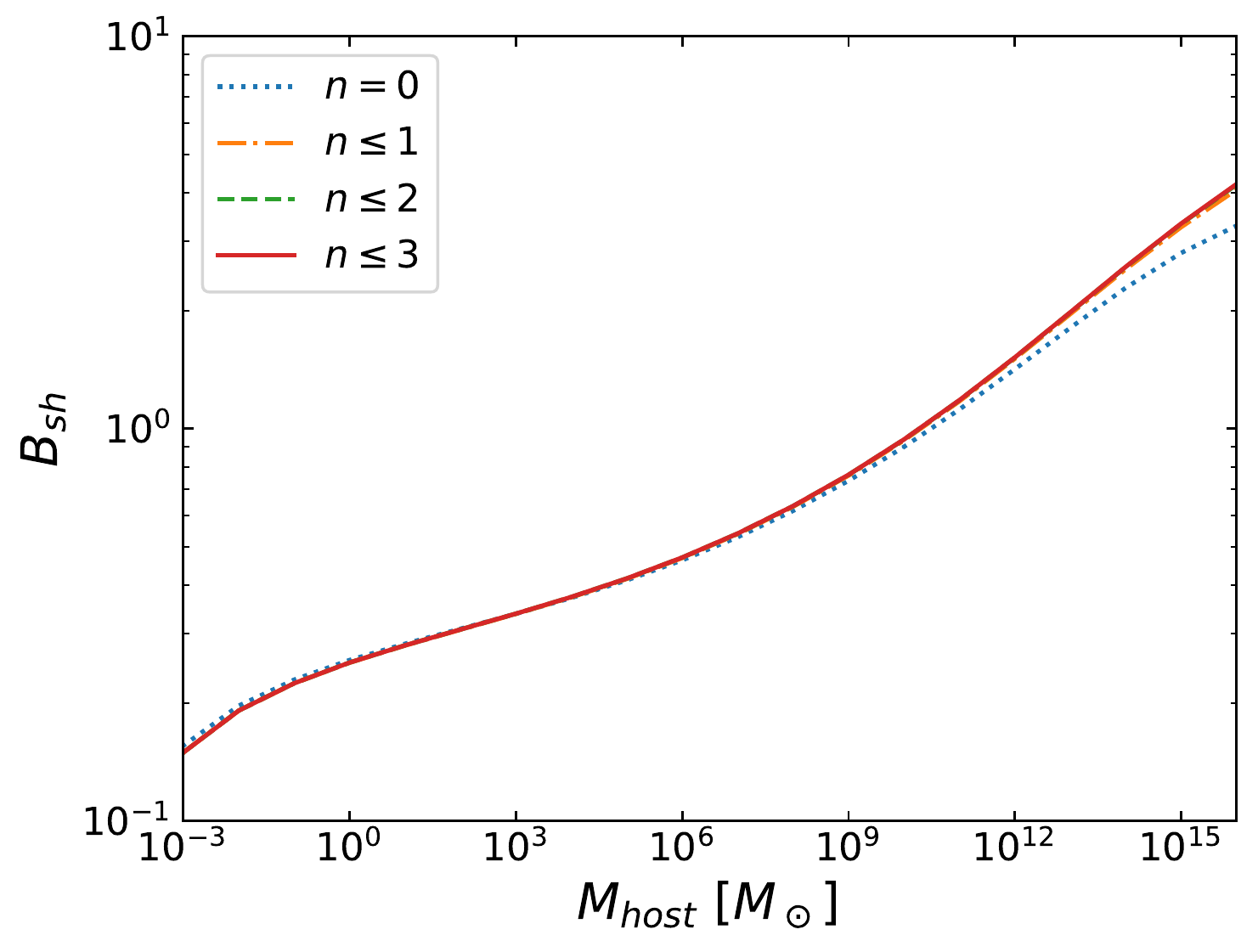}
     \includegraphics[width=7.5cm]{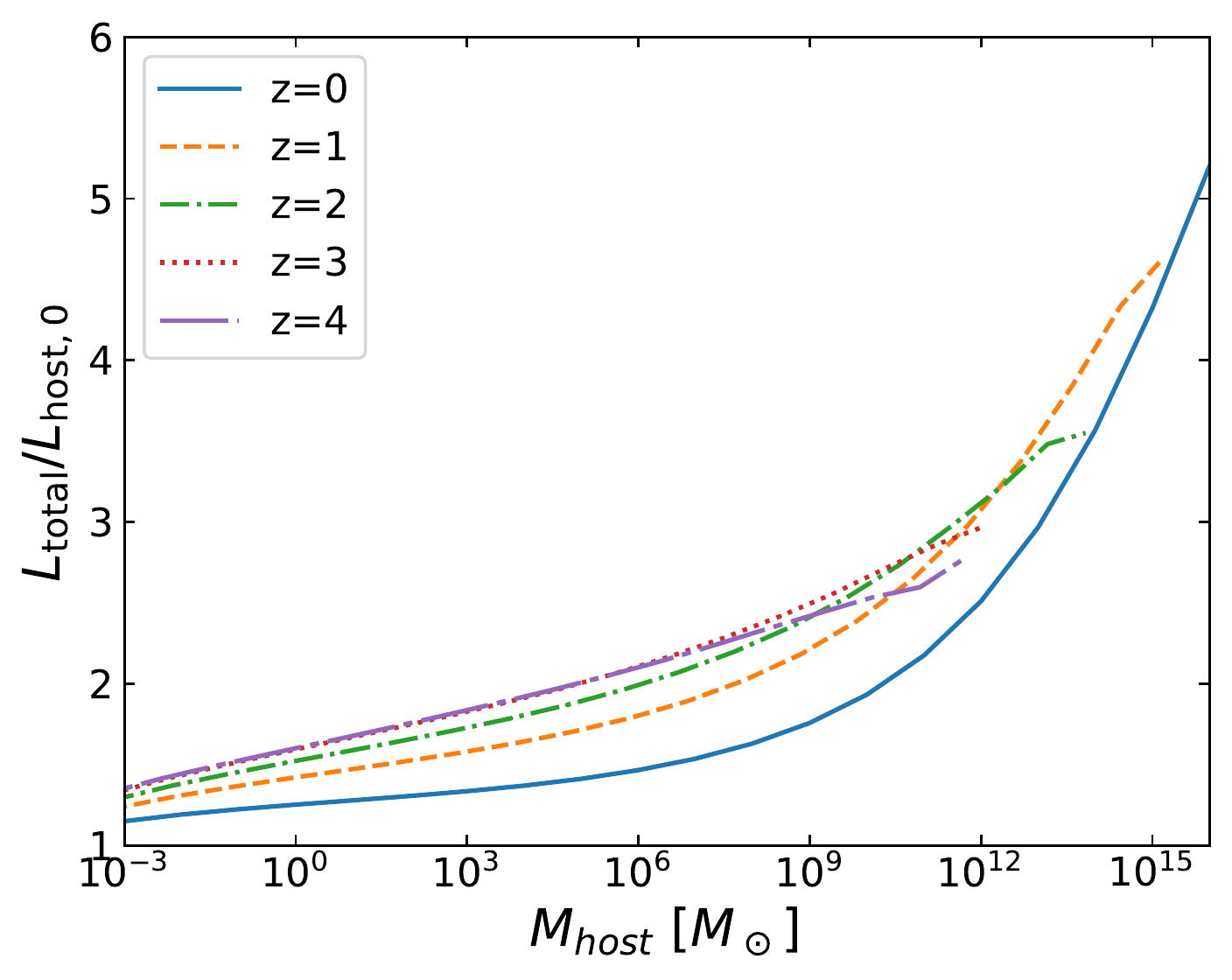}
     \includegraphics[width=7.5cm]{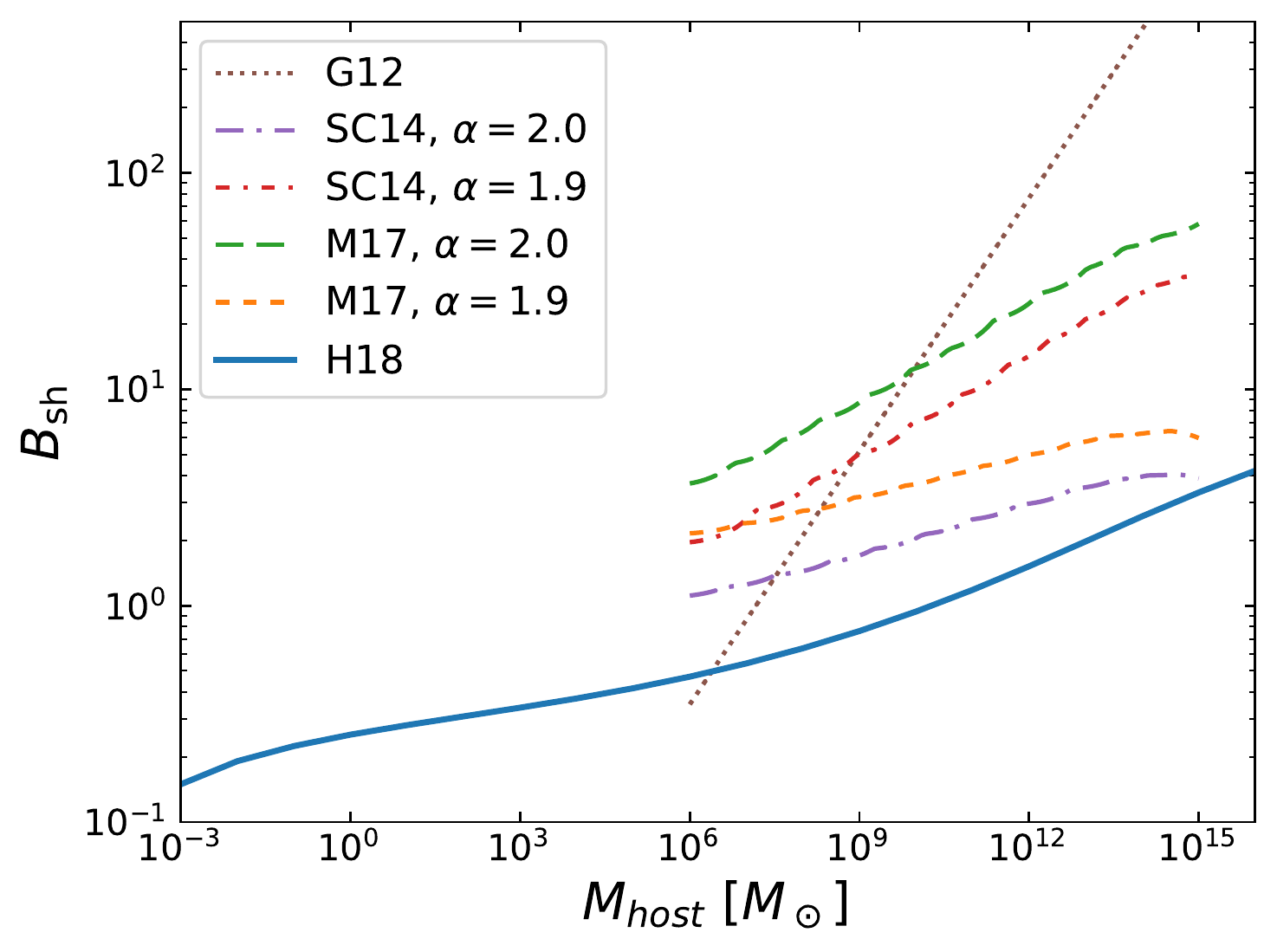}
   \caption{The subhalo boost factor $B_{\rm sh}$ as a function of the host mass $M_{200}$ for various values of redshift $z$ (top left) {based on the analytic models by \citet{Hiroshima:2018kfv}}. The~effect of sub$^{n}$-subhalos, up~to $n = 3$, is shown in the right panel in the case of $z = 0$. {Note that the three curves except for $n=0$ overlap with each other.} The bottom {left} panel shows the ratio between the total luminosity including the subhalo boost and the luminosity in absence of subhalos, $L_{\rm total}/L_{\rm host,0} = 1-f_{\rm sh}^2+B_{\rm sh}$. {The bottom right panel shows comparison of $B_{\rm sh}$ between several models at $z=0$: G12~\cite{Gao:2011rf}, SC14~\cite{Sanchez-Conde:2013yxa} and~M17~\cite{Moline:2016pbm} are based on $N$-body calculations while H18~\cite{Hiroshima:2018kfv} is on analytic calculations. The~subhalo mass function for the $N$-body results is assumed to be $dN_{\rm sh}/dm\propto m^{-\alpha}$.}}
    \label{fig:boost}
\end{figure}

{Figure~\ref{fig:boost} shows} the subhalo boost factors $B_{\rm sh}$ as a function of the host mass $M$ at various redshifts $z$ (top left).
The boost factors are on the order of unity, while it can be as larger as $\sim$5 for cluster-size halos.
It is also noted that they are larger at higher redshifts, because~the subhalos have less time to be disrupted.
The top right panel of Figure~\ref{fig:boost} shows the effect of sub$^n$-subhalos, {which} is saturated after the second iteration. 
The contribution to the boost factors due to sub-subhalos and beyond is $\lesssim$10\% for the hosts with $M_{\rm host}\geq10^{13}M_\odot$.
{The bottom left panel of Figure~\ref{fig:boost} shows} the luminosity ratio $L_{\rm total}/L_{\rm host,0} = 1-f_{\rm sh}^2 + B_{\rm sh}$ (Equation~\ref{eq:Ltotal}) as a function of the host masses for various values of the redshifts. 
{The bottom right panel of Figure~\ref{fig:boost} shows comparison with the results of the other work~\cite{Gao:2011rf, Sanchez-Conde:2013yxa, Moline:2016pbm}. We note that the analytic models do not rely on the subhalo mass function prepared separately, as~the models can provide them in a self-consistent manner. The~resulting boost factors are, however, found to be more modest than the previous results. This is mainly because the subhalo mass function adopted in the literature is larger than the predictions of the analytic models. However, they might be larger because of halo-to-halo variance. See discrepancy between predictions of the subhalo mass function for the $1.8\times 10^{12}M_\odot$ halo by \citet{Hiroshima:2018kfv} and the result of \citet{Springel:2008cc} shown in the top left panel of Figure~\ref{fig:mass function}.}

Finally, for~convenience of the reader who might be interested in using the results without going into details of the formalism, we provide fitting functions for both the subhalo mass functions and the annihilation boost factors.
They are summarized in Appendix~\ref{app:Fitting formula}.




{\subsection{Models for Self-Similar~Subhalos}
\label{sub:Models for self-similar subhalos}}

Assuming a self-similarity of the subhalos, \citet{Kamionkowski:2008vw} developed a fully analytic formulation for the probability distribution function of the dark matter density, $P(\rho)$.
Then \citet{Kamionkowski:2010mi} applied the formulation to the result of cosmological $N$-body simulations, to~obtain the fitting function for the Galactic local boost factor at Galactocentric radius $r$:
\begin{equation}
    B_{\rm sh}(r) = f_{\rm sm}(r) e^{\delta_f^2}+\left[1-f_{\rm sm}(r)\right]\frac{1+\alpha_K}{1-\alpha_K}\left[\left(\frac{\rho_{\rm max}}{\rho_\chi(r)}\right)^{1-\alpha_K}-1\right],
\end{equation}
where $f_{\rm sm}(r)$ is the volume fraction occupied by the smooth component and~$\rho_{\rm max}$ is the highest dark matter subhalo density.
Through the calibration with the numerical simulations, \mbox{\citet{Kamionkowski:2010mi}} found $\delta_f = 0.2$, $\alpha_K = 0$ and~that the subhalo fraction was given by
\begin{equation}
    1-f_{\rm sm}(r) = \kappa \left[\frac{\rho_\chi(r)}{\rho_\chi(100~{\rm kpc})}\right]^{-0.26},
\end{equation}
where $\kappa = 0.007$.
\citet{Fornasa:2012gu} then suggested a larger value of $\kappa = 0.15$--0.2 to obtain a larger boost consistent with earlier work~\cite{Springel:2008by, Pinzke:2011ek,Gao:2011rf}.
The maximum subhalo density $\rho_{\rm max}$ is estimated as $\rho_{\rm max} = [c^3/f(c)/12]200 \rho_c(z_f)$, where $c$ and $z_f$ are the concentration parameter and collapse redshift of the smallest halos.
\citet{Kamionkowski:2010mi} adopted $c = 3.5$ and $z_f = 40$ and~$\rho_{\rm max} = 80$~GeV~cm$^{-3}$.
On the other hand, \citet{Ng:2013xha} obtained a smaller $\rho_{\rm max} \approx 20$~GeV~cm$^{-3}$ even with a very small cutoff masses of $M_{\rm min} = 10^{-12} M_\odot$
Within the virial radius of the Milky-Way halo, the~subhalo boost factor for dark matter annihilation is found to be no greater than $\sim$10~\cite{Kamionkowski:2010mi}.

{\subsection{Universal Clustering of Dark Matter in Phase~Space}
\label{sub:Universal clustering of dark matter in phase space}}

\citet{Zavala:2015ura} investigated the behavior of dark matter particles that belong to the halo substructure in the phase space of distance and velocity.
Reconstructing the phase-space distribution using the Aquarius numerical simulations~\cite{Springel:2008cc}, they found universality of the coarse-grained phase-space distribution ranging from dwarfs to clusters of galaxies.
They developed physically motivated models based on the stable clustering hypothesis, spherical collapses and~the tidal stripping of the subhalos and~applied to the obtained phase-space distribution data from the simulations to find a good agreement.
Then, they computed the nonlinear matter power spectrum based on the halo model~\cite{Seljak:2000gq} down to very small free-streaming cutoff scales.
Based on the power spectrum, they obtained the subhalo boost factor greater than $\sim$30--100 for the Milky-Way size halos, which is significantly larger than the values obtained with other analytic work~\cite{Kamionkowski:2010mi, Bartels:2015uba, Hiroshima:2018kfv}. This discrepancy might come from the treatment at very small scales, where it is very hard to calibrate the analytic models against the results of the numerical~simulations.

\section{Conclusions}
\label{sec:Conclusions}

It is established that dark matter halos are made up with lots of substructures.
Especially in the cold dark matter scenario, small structures form first and~merge and accrete to create larger halos.
If~the dark matter is made of weakly interacting massive particles such as the supersymmetric neutralino, the~smallest halos can be as light as or even lighter than the Earth.
The rate of dark matter annihilation and hence its signatures such as gamma-ray fluxes are proportional to dark matter density squared and~therefore, having small-scale ``clumpy'' subhalos will \textit{boost} the~signals.

It is, however, of~an extraordinary challenge to estimate this subhalo boost factor, that is,~the ratio of luminosity from dark matter annihilation in the subhalos to that in the smoothly distributed main component.
This is mainly because subhalos of all the mass scales ranging from Earth to galaxy masses can contribute to the boost factor nearly equally per decade in mass.
In this review, we cover recent progress to overcome this issue to obtain realistic and unbiased estimates on the subhalo boost factor that will impact on interpretation of the measurements on particle physics parameters such as the annihilation cross section.
While cosmological $N$-body simulations provide the most accurate avenue to study structures in highly nonlinear regime, it is inevitably limited by the numerical resolution.
Even the state-of-the-art $N$-body simulations~\cite{Springel:2008cc, Diemand:2006ey} can resolve subhalos ranging for only several decades, which is still more than ten orders of magnitude in short to resolve all the subhalos.
Therefore, the~boost estimates have to rely on extrapolation of the subhalo properties such as its mass function and concentration parameter, which are often well described with power-law functions. 
Danger of extrapolating trends found in resolved regime for other many orders of magnitude had been widely acknowledged but~nevertheless, it was found that the estimates based on such extrapolations tended to give very large amount of boost factor of $\sim$100 ($\sim$1000) for galaxy (cluster) size halos~\cite{Springel:2008by, Gao:2011rf}.

As a complementary approach, analytic models have been investigated.
They are based on self similar propertiese of the subhalos~\cite{Kamionkowski:2008vw, Kamionkowski:2010mi}, universal phase-space distribution~\cite{Zavala:2015ura} and~extended Press-Schechter formalism combined with tidal stripping modeling~\cite{Bartels:2015uba, Hiroshima:2018kfv}.
(More recent numerical approach also adopts the concentration-mass relation calibrated for the subhalos in order to take the tidal effects into account~\cite{Moline:2016pbm}.)
Most importantly, these are all calibrated with the cosmological $N$-body simulations at resolved regimes and~proven to reproduce the simulation results such as the subhalo mass functions.
For example, the~most recent analytic models by Reference \citet{Hiroshima:2018kfv} predict the subhalo mass functions for various host masses and redshifts, which are found to be in good agreement with the simulation results~\cite{Springel:2008cc, Diemand:2006ey, Giocoli:2009ie, Ishiyama:2014gla}.
The annihilation boost factors based on these analytic models tend to be more modest, $\mathcal O(1)$ for galaxy-size halos and $\lesssim \mathcal O(10)$ for cluster-size halos.
However, none of these models have been tested against simulations at very small host halos that are less massive than $10^6 M_\odot$.
Simulations of microhalos with $10^{-6}M_\odot$ suggest cuspier profiles towards halo centers such as $r^{-1.5}$~\cite{Ishiyama:2014uoa} and~if this is the case for the subhalos too, it would boost the annihilation rate~further.

{It is known that including baryons in the simulations affects properties of subhalos such as spatial distribution and density profiles (e.g., References~\cite{Zhu:2015jwa,2017MNRAS.465L..59E, 2017MNRAS.472.4343C, Kelley:2018pdy}) and~hence there might be some effect on the annihilation boost factors.
This, however, remains largely unexplored and has to wait for future progress.
However, since the subhalos of all masses ranging down to about the Earth mass contribute to the boost factors and~the baryons will likely affect only halos of dwarf galaxies or larger, we anticipate that it is not a very important effect for the annihilation boost factors.}

The subhalo boosts directly impact the obtained upper limits on the dark matter annihilation cross section from the extragalactic halo observations.
Therefore, to~obtain the most accurate estimates of the boost factor by reducing uncertainties on structure formation at small scales as well as the physics of tidal stripping is of extreme importance for the indirect searches for particle dark matter through self-annihilation with the current and near future observations of high-energy gamma-rays, neutrinos and~charged cosmic~rays.



\authorcontributions{SA wrote most of the review sections except for Section~\ref{sec:sim}, which was contributed by TI including Figures~\ref{fig:dm_distribution}–\ref{fig:mh_boost}.  NH prepared Figures~\ref{fig:mass function} and \ref{fig:boost} and wrote Appendix~\ref{app:Fitting formula}.} 

\acknowledgments{This work was supported by JSPS KAKENHI Grant Numbers JP17H04836 (SA), JP18H04340 (SA and NH), JP18H04578 (SA), JP15H01030 (TI), JP17H04828 (TI), JP17H01101 (TI) and JP18H04337 (TI). TI has been supported by MEXT as ``Priority Issue on Post-K computer'' (Elucidation of the Fundamental Laws and Evolution of the Universe) and JICFuS. Numerical computations were partially carried out on the K computer at the RIKEN Advanced Institute for Computational Science (Proposal numbers hp150226, hp160212, hp170231, hp180180), and~Aterui supercomputer at Center for Computational Astrophysics, CfCA, of~National Astronomical Observatory of Japan.}


\conflictsofinterest{The authors declare no conflict of interest.} 

\appendixtitles{yes} 
\appendixsections{multiple} 
\appendix
\section{Fitting~Formulae}
\label{app:Fitting formula}

In this appendix, we provide fitting functions obtained with the analytical calculation in \citet{Hiroshima:2018kfv} that covers more than twenty orders of magnitude in the mass range and the redshift up to $\sim$10. The~mass function of the subhalo $m^2dN_{\rm sh}/dm$, the~luminosity of the subhalo separated from the particle physics factors $L_{\rm sh}$, and~the boost factor $B_{\rm sh}=L_{\rm sh}/L_{\rm host,0}$ [see~Equation~(\ref{eq:boost simple})] as functions of the host mass and the redshift are provided here. {We also summarize fitting functions for the boost factor $B_{\rm sh}$ at $z=0$ in the literature.} Note that the host mass is always measured in units of the solar mass ($M_\odot$) in this~appendix.

\subsection{Subhalo Mass~Function}

The fitting formula is written in the follwing form:
\begin{equation}
m^2\frac{dN}{dm}=(a+bm^\alpha)\exp\left[-\left(\frac{m}{m_c}\right)^\beta\right],
\label{eq:shmf}
\end{equation}
introducing a cutoff mass $m_c$; $a$, $b$, and~$m_c$ in Equation~(\ref{eq:shmf}) are functions of the host mass and the~redshift:
\begin{eqnarray}
a(M_{\rm host},z)&=&6.0\times10^{-4}M_\mathrm{host}(z) \left(2-0.08\log_{10}\left[M_{\rm host}(z)\right]\right)^2\left(\log_{10}\left[\frac{M_{\rm host}(z)}{10^{-5}}\right]\right)^{-0.1}\\ \nonumber
&&\times(1+z), \\
b(M_{\rm host},z)&=&8.0\times10^{-5}M_{\rm host}(z)\left(2-0.08\log_{10}\left[M_{\rm host}(z)\right]\right)\left(\log_{10}\left[\frac{M_{\rm host}(z)}{10^{-5}}\right]\right)^{-0.08z} \\ \nonumber
&&\times\left(\log_{10}\left[\frac{M_{\rm host}(z)}{10^{-8}}\right]\right)^{-1}\left(\log_{10}\left[\frac{M_{\rm host}(z)}{10^{18}}\right]\right)^2, \\
m_c&=&0.05\left(1+z\right)M_\mathrm{host}(z),\\ 
\alpha&=&0.2+0.02z, \\
\beta&=&3.
\end{eqnarray}

\begin{figure}[H]
\centering
\includegraphics[width=7.5cm]{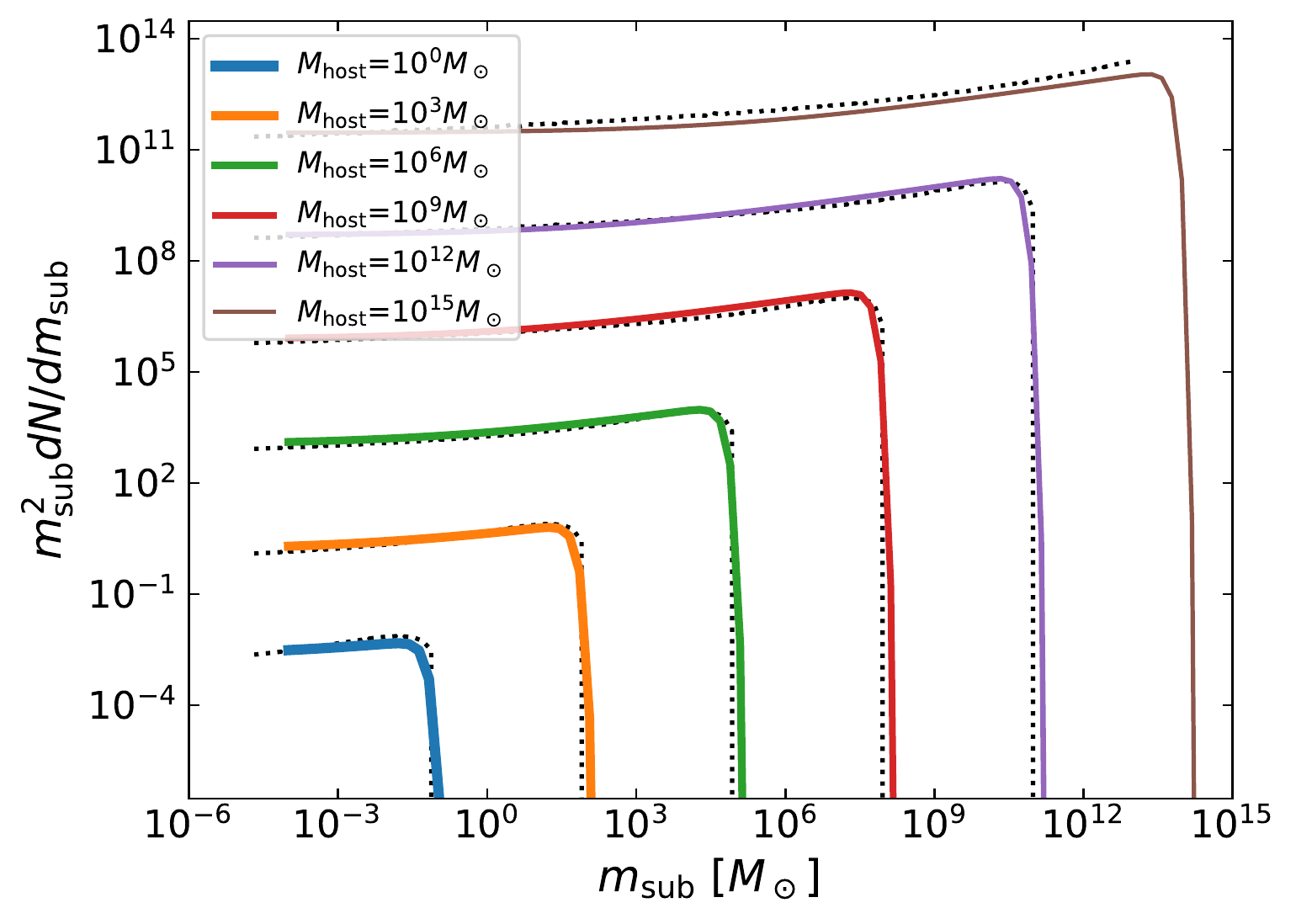}
\caption{Subhalo mass function $m^2dN/dm$ at $z=0$. Each line corresponds to a different host halo mass. The~fitting formula is applicable for the host mass from $M_{\rm host}\simeq10^{-4}$ to  $10^{14}M_\odot$ and redshifts up to $\sim$6.}
\label{f:massfcnfit}
\end{figure}

\begin{figure}[H]
\centering
\includegraphics[width=7.5cm]{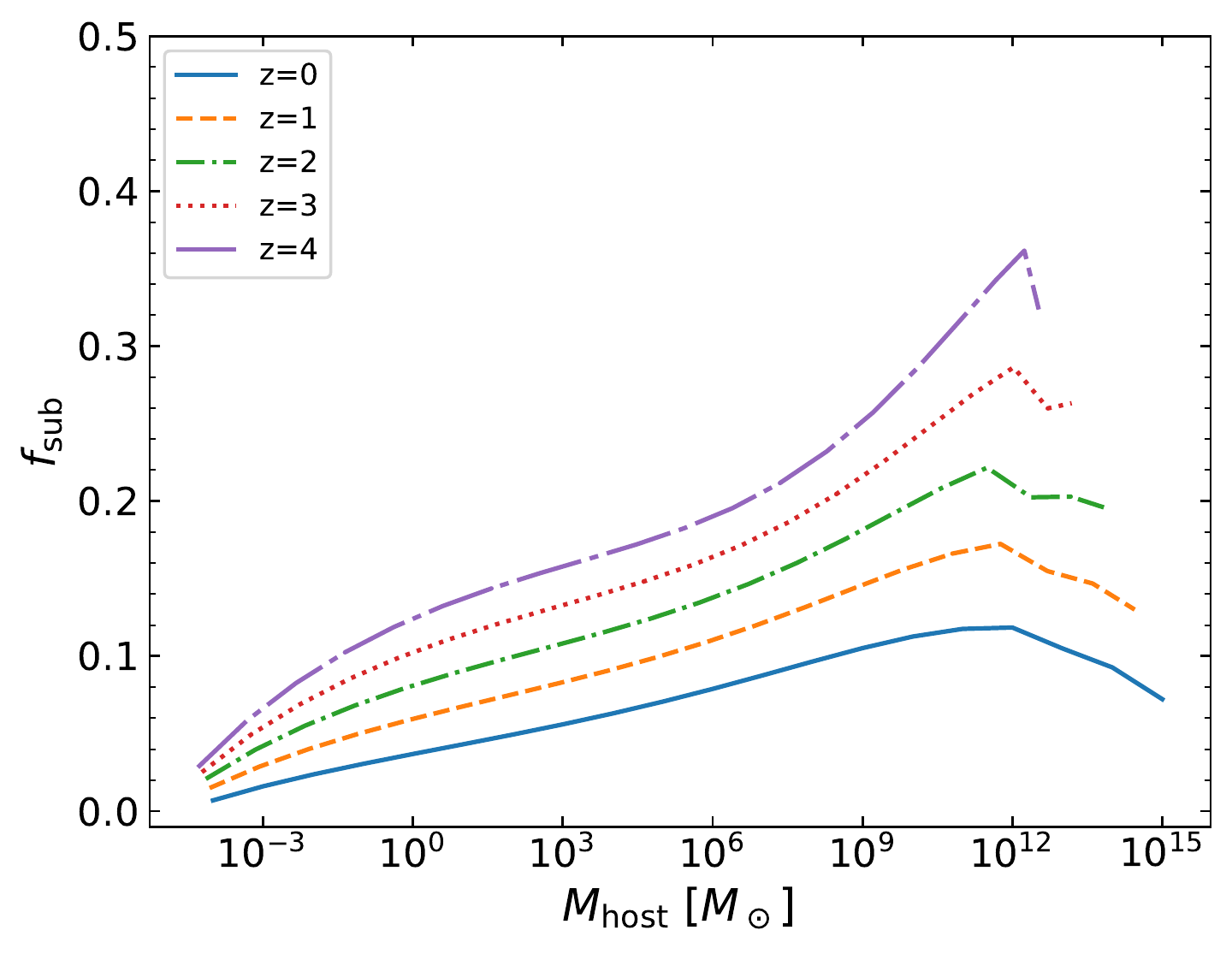}
\caption{The subhalo mass fraction obtained by integrating Equation~(\ref{eq:shmf}).}
\label{f:massffracint}
\end{figure}

In Figure~\ref{f:massfcnfit}, we show the comparison between the mass function obtained in analytical calculations~\cite{Hiroshima:2018kfv} and Equation~{(\ref{eq:shmf})}. By~integrating Equation~(\ref{eq:shmf}), we obtain the mass fraction shown in Figure~\ref{f:massffracint}.

\unskip
\unskip

\subsection{Subhalo~Luminosity}

The luminosity of a subhalo is written as
\begin{equation}
    L_{\rm host}(M)\propto\rho_s^2r_s^3\left[1-\frac{1}{(1+c_{\rm vir}^3)}\right],
\end{equation}
assuming a subhalo of NFW profile. In~this section, the~characteristic density $\rho_s$ and the scale radius $r_s$ are measured in units of g/cm$^3$ and cm, respectively. 
For~simplicity, we do not show the integration over the distribution of the virial concentration parameter $P(c_{\rm vir}|M,z)$ in the above expression. The~constant of  proportionality is detemined by fixing the particle physics model. However, we do not include it in the following expression since it cancels in the calculation of the boost factor $B_{\rm sh}$ by taking the ratio of the host and subhalo luminosity $L_{\rm sh}/L_{\rm host}$. 

The fitting formula for the luminosity takes the form of
\begin{equation}
\log_{10} L_{\rm sh}=b + a \log_{10} m,
\end{equation}
where
\begin{eqnarray}
a&=&(-0.025z+0.18)\left(\log_{10}\left[\frac{M_\mathrm{host}(z)}{10^{-5}}\right]\right)^{0.3}+(0.06z+0.53),\\
b&=&-0.95\log_{10}\left[M_{
\rm host}(z)\right]+(0.1-0.015z)\log_{10}\left[\frac{M_{\mathrm{host}}(z)}{10^{4}}\right]+0.07.
\end{eqnarray}

\subsection{Annihilation Boost~Factor}
The boost factor is sensitive to the models of the concentration-mass relation. We provide fitting functions for two different concentration-mass relation models here. One coresponds to the canonical model in Reference~\cite{Hiroshima:2018kfv} which assumes the concentration-mass relation derived in Reference~\cite{Correa:2015dva}. The~other corresponds to the concentration-mass relation in Reference~\cite{Okoli:2015dta}. For~both cases, the~fitting function of the boost factor is written in a combination of two sigmoid functions, $f(x)=(1+e^{-ax})^{-1}$,
\begin{equation}
\log_{10}B_{\rm sh}=\frac{X(z)}{1+e^{-a(z)(\log_{10}[M_{\rm host}]-m_1(z))}}+c(z)\left(1+\frac{Y(z)}{1+e^{-b(z)(\log_{10}[M_{\rm host}]-m_2(z))}}\right).
\end{equation}
Funcitons $X$, $Y$, $a$, $b$, $c$, $m_1$, and~$m_2$ depend on the redshift but~they do not on the host~mass.
\begin{itemize}
\item For Correa's concentration~\cite{Correa:2015dva}
\begin{eqnarray}
X(z)&=&2.7e^{-0.2z}+0.15,\\
Y(z)&=&0.4+\left(-0.224z+0.56\right)e^{-0.8z}, \\
a(z)&=&0.10+0.095e^{-0.5z}, \\
b(z)&=&0.03z^2-0.08z-0.83, \\
c(z)&=&0.004z^2-0.04z-0.6, \\
m_1&=&-3.17z+17.4, \\
m_2&=&-\left(0.2z-1\right)^5-4.
\end{eqnarray}
\item For Okoli's concentration~\cite{Okoli:2015dta}
\begin{eqnarray}
X(z)&=&2.2e^{-0.75z}+0.67, \\
Y(z)&=&2.5e^{-0.005z}+0.8, \\
a(z)&=&0.1e^{-0.5z}+0.22, \\
b(z)&=&0.8e^{-0.5\left(z-12\right)^4}-0.24, \\
c(z)&=&-0.0005z^3-0.032z^2+0.28z-1.12, \\
m_1&=&-2.6z+8.2, \\
m_2&=&0.1e^{-3z}-12. \\
\end{eqnarray}
\end{itemize}
All of these formulae are applicable for hosts at arbitrary redshifts up to $z\sim7$.
\begin{figure}[H]
    \centering
     \includegraphics[width=7.5cm]{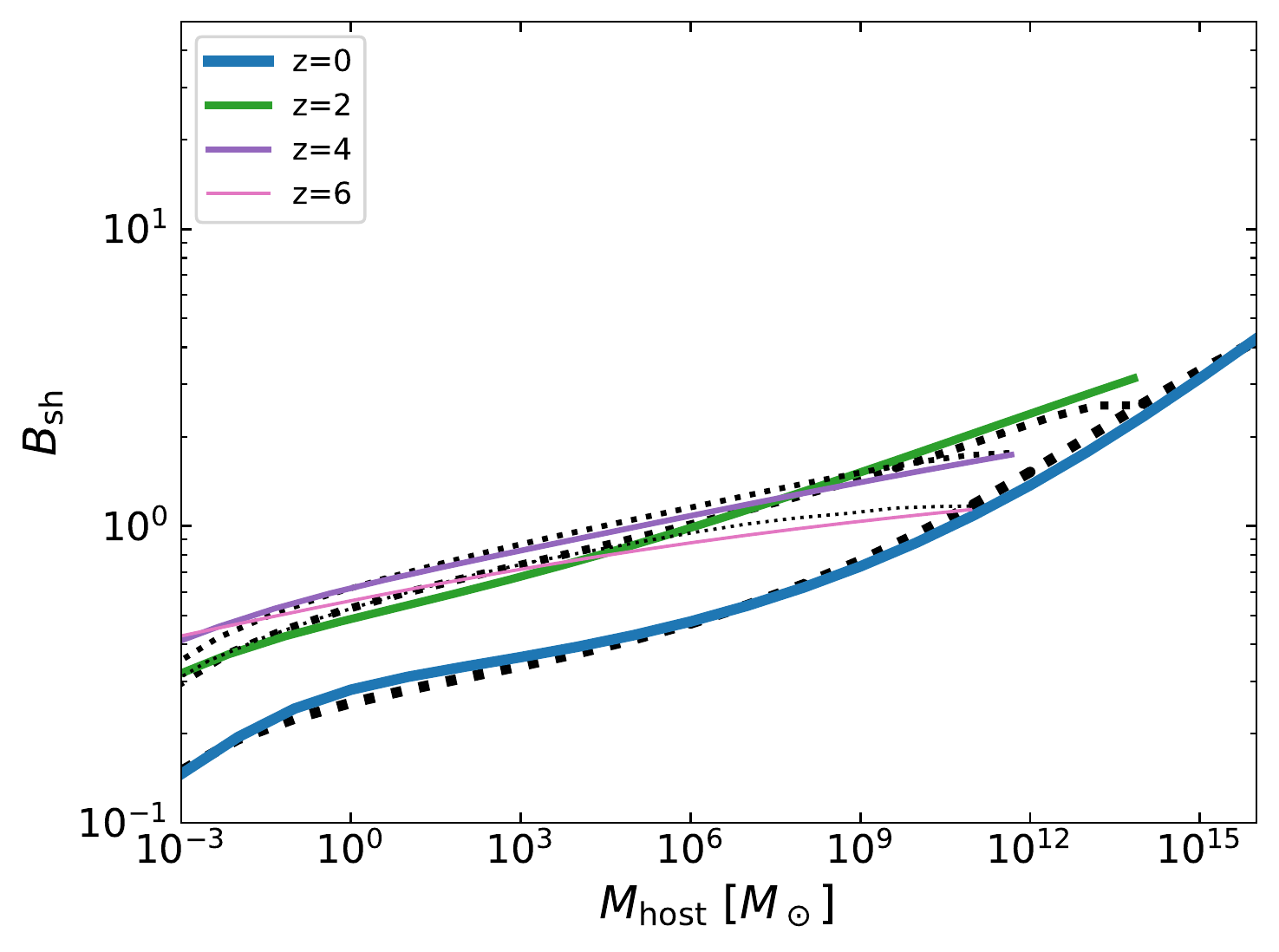}
     \includegraphics[width=7.5cm]{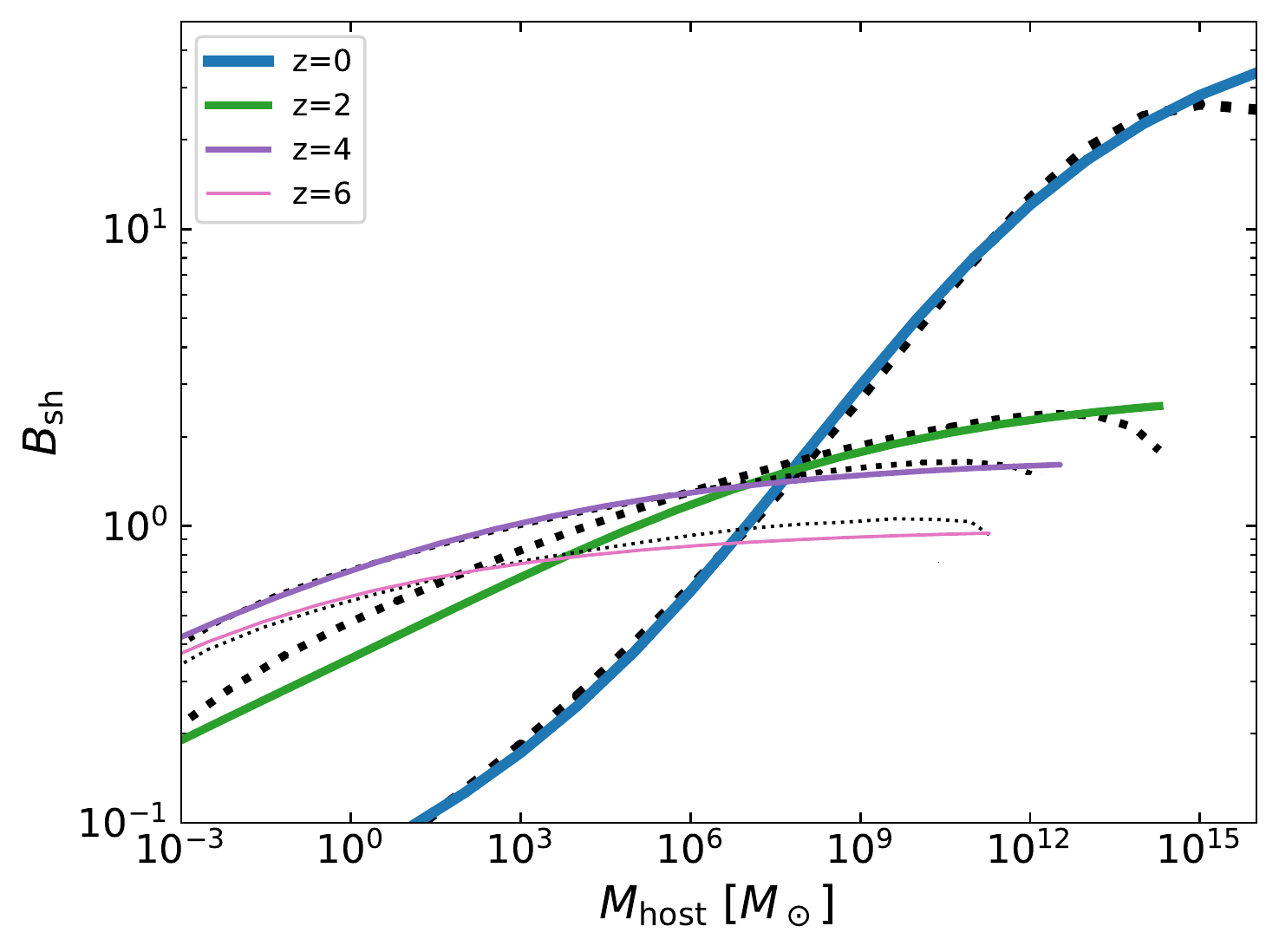}

   \caption{Comparisons between the boost factor from our calculations in Reference~\cite{Hiroshima:2018kfv} and the fitting functions in this section. The~left panel is the result assuming the concentration-mass relation in Reference~\cite{Correa:2015dva} while the right panel assuming the relation in Reference~\cite{Okoli:2015dta}.}
   \end{figure}
\unskip
   
{\subsection{Fitting Functions for the Boost Factor in the~Literature}}

Several works provide the fitting function for the boost factor at $z=0$. We summarize functions provided in \citet{Gao:2011rf,Sanchez-Conde:2013yxa}, and~\citet{Moline:2016pbm} here.

\vspace{6pt}
\begin{itemize}
    \item 
    \citet{Gao:2011rf} have analyzed cluster scale halo of $M_{\rm host}=[5,20]\times10^{14}h^{-1}M_\odot$ in the Phoenix project~\cite{Gao:2012tc}. Subhalos down to $m\sim10^{6}$ can be resolved in their calculations.
\begin{equation}
        B_{\rm sh}=1.6\times10^{-3}\left(\frac{M_{\rm host}}{M_\odot}\right)^{0.39}.
    \end{equation}
    \item 
    \citet{Sanchez-Conde:2013yxa} derive the boost factor based on the concentration-mass relation in Reference~\cite{Prada:2011jf}. The~fitting function is provided for their fiducial model assuming the minimum halo mass to be $M_{\rm min}=10^{-6}M_\odot$ and the subhalo mass function $dN/dm\propto m^{-2}$. Each subhalo is assumed to be a field halo.
\begin{equation}
        \log_{10}B_{\rm sh}(z=0)=\sum^5_{i=0}b_i\left(\ln\frac{M_{\rm host}}{M_\odot}\right)^i
        \label{fitSC}
    \end{equation}
    with
\begin{eqnarray}
    b_0&=&-0.442\\
    b_1&=&0.0796\\
    b_2&=&-0.0025\\
    b_3&=&4.77\times10^{-6}\\
    b_4&=&4.77\times10^{-6}\\
    b_5&=&-9.69\times10^{-8}
    \end{eqnarray}
    \item \citet{Moline:2016pbm} derive the boost factor taking the dependence of the survived halo properties on the distance from the host, that is,~the host potential. The~form of the function is similar to that of \citet{Sanchez-Conde:2013yxa} but adopitng different log bases:
\begin{equation}
        \log_{10}B_{\rm sh}(z=0)=\sum^5_{i=0}b_i\left(\log_{10}\frac{M_{\rm host}}{M_\odot}\right)^i
        \label{fitMoline}
    \end{equation}
    and parameters $b_i$ are
\begin{eqnarray}
    b_0&=&-0.186\\
    b_1&=&0.144\\
    b_2&=&-8.8\times10^{-3}\\
    b_3&=&1.13\times10^{-3}\\
    b_4&=&-3.7\times10^{-5}\\
    b_5&=&-2\times10^{-7}
    \end{eqnarray}
    for $\alpha=2$ and
\begin{eqnarray}
    b_0&=&-6.8\times10^{-2}\\
    b_1&=&9.4\times10^{-2}\\
    b_2&=&-9.8\times10^{-3}\\
    b_3&=&1.05\times10^{-3}\\
    b_4&=&-3.4\times10^{-5}\\
    b_5&=&-2\times10^{-7}
    \end{eqnarray}
    for $\alpha=1.9$.
\end{itemize}


\reftitle{References}



\end{document}